\documentclass[aps,rpl]{revtex4}
\usepackage{graphics,epsfig,amsmath}

\begin{document}

\title{Dynamical Mean Field Theory, Model Hamiltonians and First Principles
Electronic Structure Calculations}
\author{G. Kotliar and S. Y. Savrasov}
\address{Department of Physics and Astronomy and Center for Condensed Matter Theory,\\
Rutgers University, Piscataway, NJ 08854--8019 \\
Published in New Theoretical Approaches to Strongly Correlated Systems\\ A.
M. Tsvelik Ed.,  Kluwer Academic Publishers 259-301, (2001).}

\begin{abstract}
We review the basic ideas of the dynamical mean field theory (DMFT)
and some
of the insights into the electronic structure of strongly correlated
electrons obtained by this method in the context of model Hamiltonians. We
then discuss the perspectives for carrying out more realistic DMFT studies
of strongly correlated electron systems and we compare it with existent
methods, LDA and LDA+U. We stress the existence of new functionals for
electronic structure calculations which allow us to treat situations where
the single--particle description breaks down such as the vicinity of the
Mott transition.
\end{abstract}

\pacs{71.20.-b, 71.27.+a, 75.30.-m}

\maketitle



\section{Introduction}

The last two decades have witnessed a revival in the study of strongly
correlated electron systems. A large variety of transition metal compounds,
rare earth and actinide based materials have been synthesized. Strong
correlation effects are also seen in organic metals, and carbon based
compounds such as Bucky balls and carbon nanotubes. These systems display a
wide range of physical properties such as high--temperature
superconductivity, heavy--fermion behavior, and colossal magnetoresistance
to name a few \cite{imada}.

Strong correlation effects are the result of competing interactions. They
often produce at low temperatures several thermodynamic phases which are
very close in free energy, resulting in complex phase diagrams. As a result
of these competing tendencies, strongly--correlated electron systems are
very sensitive to small changes in external parameters, i.e. pressure,
temperature, composition, stress. This view is supported by a large body of
experimental data as well as numerous controlled studies of various models
of strongly--correlated electron systems \cite{imada,review}.

At the heart of the strong--correlation problem is the competition between
localization and delocalization, i.e. between the kinetic energy and the
electron--electron interactions.
When the overlap of the electrons among themselves is large, a wave--like
description of the electron is natural and sufficient. Fermi--liquid theory
explains why in a wide range of energies systems, such as alkali and noble
metals, behave as weakly interacting fermions, i.e. they have a Fermi
surface, linear specific heat and a constant magnetic susceptibility and
charge compressibility. The one--electron spectra form quasi--particle and
quasi--hole bands. The transport properties are well described by the
Boltzmann theory applied to long lived quasi--particles, the approach that
makes sense as long as $k_{f}l\gg 1$. Density functional theory (DFT)\ in
the local density or generalized gradient approximations (LDA or GGA), is
able to predict most physical properties with remarkable accuracy. \cite
{DFTbook}

When the electrons are very far apart, a real--space description becomes
valid. A solid is viewed as a regular array of atoms where each atom binds
an integer number of electrons. These atoms carry spin and orbital quantum
numbers giving rise to a natural spin and orbital degeneracy. Transport
occurs via activation, namely the creation of vacancies and doubly occupied
sites. Atomic physics calculations together with perturbation theory around
the atomic limit allows us to derive accurate spin--orbital Hamiltonians.
The one--electron spectrum of the Mott insulators is composed of atomic
excitations which are broadened to form bands that have no single--particle
character, known as Hubbard bands. In large number of compounds spin and
orbital degrees of freedom order at low temperatures breaking spin rotation
and spatial symmetries. However, when quantum fluctuations are strong enough
to prevent the ordering, possible new forms of quantum mechanical ground
states may emerge\cite{PWA}.

These two limits, well separated atoms, and well overlapping bands, are by
now well understood and form the basis of the ''standard model'' of
solid--state physics. One of the frontiers in strongly correlated electron
physics problem is the description of the electronic structure of solids
away from these limits. The challenge is to develop new concepts and new
computational methods capable to describing situations where both itineracy
and localization are simultaneously important. The ''standard model'' of
solids breaks down in this situation, and strongly correlated electron
systems have many anomalous properties, such as resistivities which far
exceed the Ioffe--Regel--Mott limit ${\rho _{Mott}}^{-1}\approx ({%
e^{2}/h)k_{f}}$\cite{kivelson}, non--Drude--like optical conductivities, and spectral
functions which are not well described by the band theory \cite{imada}.

To treat these systems one needs a technique which is able to describe
Kohn--Sham bands and Hubbard bands on the same footing, and which is able to
interpolate between the atomic and the band limit. Dynamical mean--field
theory (DMFT)\cite{review} is the simplest approach satisfying these
requirements. We introduce it in a very general formulation, the cellular
DMFT or C--DMFT\cite{CDMFT}, which is particularly well suited for
electronic--structure calculations in section II .

The goal of these lectures is to introduce recent DMFT developments to both
the electronic--structure community as well as researchers interested in the
many--body physics of correlated materials. For the electronic--structure
community, DMFT is a promising technique for going beyond the LDA method. To
illustrate the promise of the technique we describe in section III some of
striking progress which has been achieved in the theory of the Mott
transition by the use of DMFT at the level of model hamiltonians.

The view of strongly--correlated electron systems that we describe in this
introduction, stresses the need for incorporating electronic--structure
methods in treating strongly correlated electron systems. The {\it low
temperature} physics of systems near localization--delocalization crossover,
is non universal, system specific, and very sensitive to the lattice
structure and orbital degeneracy which are unique to each compound. We
believe that incorporating this information into the many--body treatment of
these systems is a necessary first step before more general lessons about
strong--correlation phenomena can be drawn. The extreme sensitivity of the
materials properties to microscopic details has motivated us to realistic
studies of correlated solids within DMFT. To put these efforts in an
electronic--structure perspective, and to stress the qualitative difference
between DMFT and other electronic structure methods, we review in sections
IV and V the density functional method and the LDA + U method from an
effective action point of view. In the following sections VI--XII we rely
heavily on the effective--action formulation of dynamical mean--field theory
for electronic--structure calculations \cite{chitrafunc1} \cite{chitrafunc2}

In section X\label{results} we argue that many DMFT results obtained so far
are in much better agreement with experiments than the corresponding results
of LDA calculations We conclude in section XII\label{edmft} with a brief
introduction to the E--DMFT method, a complementary approach to C--DMFT
which can take into account the longer range of the Coulomb interactions.
This method, which we call the GWU method generalizes both the well
established GW method \cite{GWreview}, and the DMFT but has not yet been
implemented \ in a realistic framework.

\section{Cellular DMFT}

\label{dmft}Reference\onlinecite{review} reviews the pre 1995 work on the
dynamical mean--field method and its various extensions. In this section we
discuss a recently proposed \cite{CDMFT} cellular version of the theory  
or C--DMFT.  This formulation which is well suited for electronic--structure
calculations since it is adapted to a non--orthogonal basis. This supercell
(or cluster) DMFT remains close in spirit to the DMFT ideas, where the
clusters have free (and not periodic) boundary conditions.\ Furthermore, the
flexibility of this approach stresses the connection between the lattice
many--body problem and self--consistent impurity models as in the single
site dynamical mean--field theory \cite{antoine} \ The construction is
carried out in complete analogy with the standard dynamical mean--field
construction \cite{review}, but allows the use of a large class of basis
sets. This frees us from the need to use sharp division of space into
supercells.

It has been proved \cite{CDMFT} that the C--DMFT\ construction is manifestly
causal, i.e.~the self--energies that result from the solution of the cluster
equations obey $\Sigma ({\bf k},\omega )\leq 0$, eliminating a priori one of
the main difficulties encountered earlier in devising practical cluster
schemes.

It is useful to separate three essential elements of a general DMFT\ scheme:
(a) Definition of the cluster degrees of freedom, which are represented by
impurity degrees of freedom in a bath described by a Weiss field matrix
function $G_{0}$. The solution of the cluster embedded into a medium results
in a cluster Green's function matrix and a cluster self--energy matrix. (b)
The expression of the Weiss field in terms of the Green function or the
self--energy of the cluster, i.e.~the self--consistency condition of the
cluster scheme. (c) The connection between the cluster self--energy and the
self--energy of the lattice problem. The impurity solver estimates the local
correlations of the cluster, while the lattice self--energy is projected out
using additional information, i.e. periodicity of the original lattice.

Our construction applies to very general models for which lattice
formulation naturally appears. It can be thought of as an extension of the
band--structure formalism that takes into account the electron--electron
interactions. The lattice hamiltonian, $H[f_{i\sigma },f_{i\sigma }^{\dagger
}]$, (one example could be the well--known Hubbard hamiltonian) is expressed
in terms of annihilation and creation operators $f_{i\sigma }$ and $%
f_{i\sigma }^{\dagger }$ where $i$ runs over the sites of a $d$--dimensional
infinite lattice $i=(i_{1},\ldots ,i_{d})$, the index $\sigma $ denotes an
internal degree of freedom such as a spin index or a spin--orbital or band
index if we consider an orbitally degenerate solid.

({\it a) Selection of cluster variables:} The first step in a mean--field
approach to a physical problem, is a selection of a finite set of relevant
variables. This is done by splitting the original lattice into clusters of
size $\prod_{j=1}^{d}L_{j}$ arranged on a superlattice with translation
vectors $R_{n}$. On this superlattice we choose wave functions $|R_{n}\alpha
\rangle $ partially localized around $R_{n}$ with $\alpha =1,\ldots ,N$
denoting an internal {\it cluster} index. The relation between the new wave
functions, $|R_{n}\alpha \rangle $, and the old ones, $|i\sigma \rangle $,
is encoded in a transformation matrix, $S_{R_{n}\alpha ,i\sigma }$, such
that $|R_{n}\alpha \rangle =\sum_{i\sigma }|i\sigma \rangle S_{i\sigma
,R_{n}\alpha }^{-1}$. Due to the translation symmetry of the lattices we
have $S_{R_{n}\alpha ,i\sigma }=S_{\alpha \sigma }(r(i)-R_{n})$ where $r(i)$
is the position of site $i$. The creation and annihilation operators of the
new basis are related to the operators of the old basis by $c_{R_{n}\alpha
}=\sum_{i\sigma }S_{R_{n}\alpha ,i\sigma }f_{i\sigma }$ and the operators
that contain the "local" information that we want to focus our attention on
are $c_{\alpha }\equiv c_{(R_{n}=0)\alpha }$, i.e.~the operators of the
cluster at the origin. We will refer to these operators as the cluster
operators. Note that we do not require that the wave--function basis is
orthogonal, and the nonorthogonality is summarized in an overlap matrix $%
O_{\mu \nu }^{mn}=O_{\mu \nu }(R_{m}-R_{n})\equiv \langle R_{m}\mu |R_{n}\nu
\rangle $.

The next step is to express the hamiltonian in terms of the complete set of
operators $c_{R_{m}\mu }$. In terms of the new set of variables it has the
form 
\begin{eqnarray}
H &=&-\sum_{R_{m}\mu R_{n}\nu }t_{\mu \nu }(R_{m}-R_{n})c_{R_{m}\mu
}^{+}c_{R_{n}\nu }  \nonumber \\
&+&\sum_{R_{1}\mu R_{2}\nu R_{3}\rho R_{4}\varsigma }U_{\mu \nu \rho
\varsigma }(\{R_{i}\})c_{R_{1}\mu }^{+}c_{R_{2}\nu }^{+}c_{R_{4}\varsigma
}c_{R_{3}\rho }  \label{hubbard}
\end{eqnarray}
We stress again  the generality of the method. Equation (\ref{hubbard})
has the form one would obtain by writing the full hamiltonian of electrons
in a solid in some tight--binding non--orthogonal basis. The hamiltonian is
then split into three parts, $H=H_{c}+H_{cb}+H_{b}$ where $H_{c}$ involves
only the cluster operators and their adjoints, $H_{b}$ contains $c_{R_{n}\mu
}$ with $R_{n}\neq 0$ only and plays the role of a ''bath'', and finally $%
H_{cb}$ contains both $c_{R_{n}\mu }$ with $R\neq 0$ and the cluster
operators $c_{\mu }$ (which have $R_{n}=0$). Physically $H_{cb}$ couples the
cluster with its environment. A similar separation can be carried out at the
level of the action, in the coherent state functional integral formulation
of this problem, where the partition function and the correlation functions
are represented as averages over Grassman variables, 
\begin{equation}
Z=\int \prod_{R_{n}\alpha }Dc_{R_{n}\alpha }^{+}Dc_{R_{n}\alpha }e^{-S}
\end{equation}
where the action is given by 
\begin{eqnarray}
S &=&\int_{o}^{\beta }\!\!d\tau \left( \sum_{R_{m}\mu R_{n}\nu }\!c_{{%
R_{m}\mu }}^{+}O_{\mu \nu }^{mn}\partial _{\tau }c_{{R_{n}\nu }}+H[c_{{%
R_{m}\mu }}^{+},c_{{R_{n}\nu }}]\right)  \nonumber \\
&\equiv &S_{c}+S_{cb}+S_{b}  \label{Shubbard}
\end{eqnarray}
The effective action for the cluster degrees of freedom is obtained
conceptually by integrating out all the variables $c_{R_{n}\mu }$ with $%
R_{n}\neq 0$ in a functional integral to obtain an effective action for the
cluster variables $c_{\mu }\equiv c_{R_{m}=0\mu }$, i.e. 
\begin{equation}
{\frac{{1}}{{Z_{eff}}}}e^{-S_{eff}[c_{\mu }^{+}c_{\mu }]}\equiv {\frac{{1}}{{%
Z}}}\int \prod_{R_{m}\neq 0,\mu }Dc_{R_{m}\mu }^{+}Dc_{R_{m}\mu }e^{-S}
\label{seff}
\end{equation}
Note that the  knowledge of the exact $S_{eff}$ allows us to calculate {\it all
local} correlation functions involving cluster operators. As described in 
\cite{review}, this cavity construction if carried out exactly would
generate terms of arbitrary high order in the cluster variables. Our
approximation renormalizes the quadratic term, and neglects the
renormalization of the quartic and the generation of higher order terms.
Since the action $S_{cb}$ contains only boundary terms, the effects of these
operators will decrease as the size of the cluster increases. Within these
assumptions, the effective action is parameterized by $G_{0,\mu \nu }(\tau
-\tau ^{\prime })$, the Weiss function of the cluster and has the form

\begin{eqnarray}
S_{eff} &=&-\int_{0}^{\beta }d\tau d\tau ^{\prime }\sum_{\mu \nu }c_{\mu
}^{+}(\tau )G_{0,\mu \nu }^{-1}(\tau -\tau ^{\prime }){c_{\nu }}(\tau
^{\prime })  \label{Seff} \\
&&+\int_{0}^{\beta }d\tau _{1}d\tau _{2}d\tau _{3}d\tau _{4}\Gamma _{\mu \nu
\rho \varsigma }c_{\mu }^{+}(\tau _{1})c_{\nu }^{+}(\tau _{2})c_{\varsigma
}(\tau _{4})c_{\rho }(\tau _{3})  \nonumber
\end{eqnarray}
where $\Gamma _{\mu \nu \rho \varsigma }=U_{\mu \nu \rho \varsigma }(\{0\})$%
. Using the effective action (\ref{Seff}) one can calculate the Green
functions of the cluster $G_{c,\mu \nu }(\tau -\tau ^{\prime })[G_{0}]\equiv
-\langle T_{\tau }c_{\mu }(\tau )c_{\nu }^{+}(\tau ^{\prime })\rangle
\lbrack G_{0}]$ and the cluster self energies 
\begin{equation}
\Sigma _{c}\equiv G_{0}^{-1}-G_{c}^{-1}.
\end{equation}

({\it b) Self-consistency condition:} The cluster algorithm is fully defined
once a self--consistency condition which indicates how $G_{0}$ should be
obtained from $\Sigma _{c}$ and $G_{c}$ is defined. In the approach that we
propose here the self--consistent equations become matrix equations
expressing the Weiss field in terms of the cluster self--energy matrix $%
\Sigma _{c}$. 
\begin{equation}
G_{0}^{-1}=\left( \sum_{{\bf k}}[(i\omega +\mu )O({\bf k})-t({\bf k})-\Sigma
_{c}]^{-1}\right) ^{-1}+\Sigma _{c}.  \label{scc}
\end{equation}
where $O({\bf k})$ is the Fourier transform of the overlap matrix, $t({\bf k}%
)$, is the Fourier transform of the kinetic energy term of the hamiltonian
in Eq.~(\ref{hubbard}) and ${\bf k}$ is now a vector in the reduced
Brillouin zone (reduced by the size of the cluster, $L_{j}$, in each
direction). Equations (\ref{seff}) and (\ref{scc}) can be derived by scaling
the hopping between the supercells as the square root of the coordination
raised to the Manhattan distance between the supercells and generalizing the
cavity construction of the DMFT \cite{review} from scalar to matrix
self--energies. If the cluster is defined in real space and the self--energy
matrices could be taken to be cyclic in the cluster indices so that the
matrix equations could be diagonalized in a cluster momentum basis, Eq.(\ref
{scc}) would reduce to the DCA equation \cite{dca}. However, in the DMFT
construction, the clusters have free and not periodic boundary conditions,
and we treat a more complicated problem requiring additional matrix
inversions.

({\it c) Connection to the self--energy of the lattice:} The
self--consistent solution, $G_{c}$ and $\Sigma _{c}$, of the cluster problem
can be related to the correlation functions of the original lattice problem
through the transformation matrix $S_{R_{m}\alpha ,i\sigma }$ by the
equation 
\begin{equation}
\Sigma _{lat,\sigma \sigma ^{\prime }}(k,\omega )=\sum_{\mu \nu }\tilde{S}%
_{\sigma ,\mu }^{\dagger }(k)\Sigma _{c,\mu \nu }(\omega )\tilde{S}_{\nu
,\sigma ^{\prime }}(k)  \label{eight}
\end{equation}
where $\tilde{S}$ is the Fourier transform of the matrix S with respect to
the original lattice indices $i$. Notice that $\Sigma _{lat,\sigma \sigma
^{\prime }}$ is diagonal in momentum and will also be diagonal in the
variable $\sigma $ if this variable is conserved.

({\it d) Connection to impurity models:} As within the single--site DMFT it
is very convenient to view the cluster action as arising from a hamiltonian, 
\begin{eqnarray}
H_{imp} &=&\sum_{\rho \varsigma }E_{\rho \varsigma }c_{\rho
}^{+}c_{\varsigma }+\sum_{\mu \nu \rho \varsigma }\Gamma _{\mu \nu \rho
\varsigma }c_{\mu }^{+}c_{\nu }^{+}c_{\rho }c_{\varsigma }  \nonumber \\
&&+\sum_{{\bf k}j}\epsilon _{{\bf k}j}a_{{\bf k}j}^{+}a_{{\bf k}j}+\sum_{%
{\bf k}j,\mu }\left( V_{{\bf k}j,\mu }a_{{\bf k}j}^{+}c_{\mu }+h.c.\right) .
\end{eqnarray}
Here $\epsilon _{{\bf k}j}$ is the dispersion of the auxiliary band and $V_{%
{\bf k}j,\mu }$ are the hybridization matrix elements describing the effect
of the medium on the impurity. When the band degrees of freedom are
integrated out the effect of the medium is parameterized by a hybridization
function, 
\begin{equation}
\Delta _{\mu \nu }(i\omega _{n})[\epsilon _{{\bf k}j},V_{{\bf k}j}]=\sum_{%
{\bf k}j}\frac{V_{{\bf k}j,\mu }^{\ast }V_{{\bf k}j,\nu }}{i\omega
_{n}-\epsilon _{{\bf k}j}}.
\end{equation}
The hybridization function is related to the Weiss--field function by
expanding Eq.(\ref{scc}) in high frequencies: 
\begin{equation}
G_{0}^{-1}(i\omega _{n})=i\omega _{n}O-E-\Delta (i\omega _{n})
\label{eq:Weiss}
\end{equation}
with $O=\left[ \sum_{{\bf k}}O_{{\bf k}}^{-1}\right] ^{-1}$ indicating that
the impurity model has been written in a non--orthogonal local basis with an
overlap matrix $O$

Finally we note that one can view the C-DMFT as an alternative to the usual
approach to treating finite systems by imposing on them periodic boundary
conditions. Here we use a boundary condition which is physically equivalent
to embedding the cluster in an infinite medium, which is determined self
consistently using information computed within the cluster, and structural
information on how the cluster is embedded in the infinite lattice.

\section{Qualitative Insights from DMFT applied to model Hamiltonians}

\label{qualitative}

\subsubsection{DMFT phase diagrams, Frustration, Complexity and Universality}

The low--temperature phase diagram of simple hamiltonians treated within
DMFT has several distinct phases, and is fairly complex. Even the simplest,
bare--bones hamiltonian (one--band Hubbard model with partial frustration)
has at least a metallic antiferromagnetic phase and a paramagnetic
insulating phase in addition to a paramagnetic metal phase and the
antiferromagnetic insulating phase.
The understanding of this model required several years of research
efforts by several groups \cite{prls} \cite{mj} \cite{zrk} \cite{werner}
\cite{rozenberg} \cite{lange} \cite{landau}. In this section we highlight some of the insights
obtained, to motivate the need  for extending the  DMFT  method to
incorporate  realistic aspects of the electronic structure.

\begin{figure}[htb]
\center{\epsfig{file=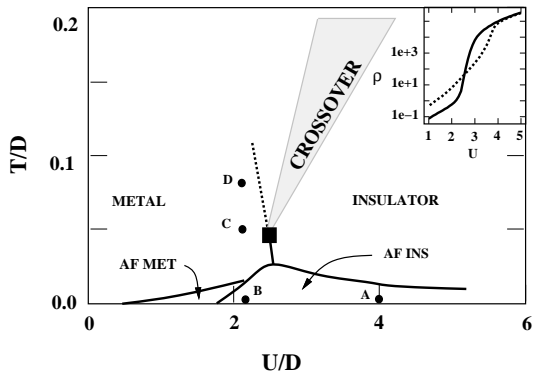,height=50mm,angle=0}}
\caption{Schematic phase diagram of partially frustrated Hubbard model from
ref \protect\onlinecite{rozenberg}, the inset illustrates the behaviour of the
resistivity above but near the Mott endpoint}
\label{fig5}
\end{figure}

The phase diagram\cite{rozenberg} shown
in Fig \ref{fig5} shares some similarities with the phase diagram of
Vanadium oxide. This observation leads Rozenberg et. al.\cite{rozenberg} to
suggest several optical experiments which have confirmed some qualitative
predictions of DMFT. These and other successful predictions of DMFT applied
to simple model of correlated electrons are described in this section to
motivate the applications of this technique in a more realistic setting.

It is important to emphasize, however, that the main lesson drawn from the
qualitative similarity in the low--temperature region between the DMFT phase
diagram of one of the simplest models of correlated electron systems and
that of some real oxides, is the ability of the DMFT method to capture
multiplicity of possible ordered states. The detailed nature of these phases
and the character of the transitions between them depend on many details of
the hamiltonian describing the specific crystal structure and chemistry of
the compound. To approach this problem realistic versions of DMFT have been
constructed and are being developed\cite{poteryaev,savrasov}.

Strong dependence of the low--temperature phases and of the low--temperature
physical properties of each material on its crystal structure and chemical
composition should be contrasted with the remarkable degree of universality
that is predicted to occur at higher temperatures. All that is required to
produce the high--temperature features of the DMFT phase diagram is a large
degree of magnetic frustration to suppress the long--range order and to
allow for a localized phase with a large entropy content. In systems without
magnetic frustration, the onset of magnetism or other forms of order
preempts us from accessing this strongly correlated regime. The origin of
the magnetic frustration is crucial for understanding the low--temperature
part of the phase diagram, with its myriad of ordered phases, but is rather
irrelevant in the high--temperature regime, where thermal fluctuations
average all various configurations leading to a more universal description
which is captured by a relatively local approach such as DMFT in its single
site or in its clusters versions. In systems such as titanates and vanadium
oxides, the origin of frustration arises from the orbital degeneracy which
is unique to those materials. In nickel selenide sulfide mixtures, the
crystal structure is such that a sizeable ring exchange term competes with
the nearest--neighbor superexchange interaction resulting in a reduced Neel
temperature. Still, these systems display very similar phenomena around the
Mott transition endpoint.

Contrast between highly--universal behavior at high temperature and the
dependence of low--temperature properties on additional parameters in the
Hamiltonian, was discussed\cite{chitraaf} in connection with the comparison
of the physical properties of the vanadium oxide and the nickel selenide
sulfide mixtures. The phase diagram of the two--dimensional organic compound 
$\kappa $ BEDTTF \cite{lefevre}, where the frustration originates in its
underlying chiral triangular lattice of dimers, strengthen the validity of
this point of view. Indeed many of the high--temperature physical properties
of this have been accounted for by the DMFT studies of McKenzie and Merino 
\cite{mackenzie}.

To summarize, since magnetic frustration and competition of kinetic and
interaction energy are all that is required for obtaining the
high--temperature part of the ``canonical'' phase diagram of a correlated
electron system. At low temperatures between two values of U , $U_{c1}$ and $%
U_{c2}$ two dynamical mean field solutions are possible. The transition
between the localized and extended regime as a function of $\frac{U}{t}$
takes place via a first--order transition \cite{werner}\cite{rozenberg},
this is faithfully reproduced by the simplest model containing these
ingredients treated within DMFT.

The phase diagram\cite{rozenberg} displays two crossover lines. The dotted
line in fig 1 is a coherence incoherence crossover (i.e. the continuation of
the $U_{c2}$ line where metallicity is lost). The shaded area is a
continuation of the $U_{c1}$ line, where the temperature becomes comparable
with the gap. Both were observed in the $V_{2}O_{3}$ and $NiSeS$ system \cite
{kuwamoto,takagi}. Further justification for this point of view, and a
refined description of the localization delocalization transition around the
Mott transition endpoint was achieved by the development of a Landau like
description \cite{landau,lange}.

\subsubsection{Coherent and Incoherent Spectra}

Mapping onto the Anderson impurity model offers an intuitive picture of both
metallic states and Mott insulating states. A correlated metal is described
locally as the Anderson impurity model in a metallic bath: The Kondo effect
gives rise to strongly renormalized quasiparticles when the interactions are
strong, and to a broad band when the interactions are weak. The Mott
insulator is locally described as the Anderson impurity model in an
insulating bath. The charge degrees of freedom are gapped, but the spin
degrees of freedom are not quenched; they dominate the low--energy physics.
When there is one electron per site, the Mott transition takes place as one
goes from the first regime to the second by increasing the strength of the
interaction $U$ \cite{prls}.

\begin{figure}[b]
\centerline{\epsfig{file=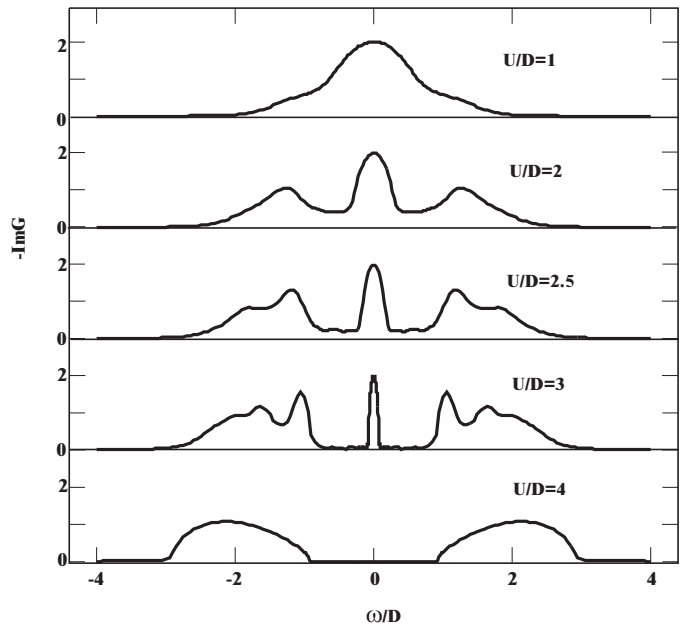,width=3.5in}}
\caption{Evolution of the spectral function at zero temperature as function
of U. {} From Ref. \protect\onlinecite{zrk}.}
\label{spectralu}
\end{figure}

A sketch of the evolution of the spectral function $-ImG(i\Omega +i\delta )$
of the half--filled Hubbard model is described in figure (\ref{spectralu}).
For interactions $U$ close but smaller than the critical $U_{c2}$ the
one--electron spectral function of the Hubbard model in the strongly
correlated metallic region contains both atomic features (i.e. Hubbard
bands) and quasiparticle features in its spectra \cite{antoine}. This may be
understood intuitively from the Anderson--Yuval--Hamman path--integral
representation. In the regime of strong correlations, paths that are nearly
constant in imaginary time, as well as those that fluctuate strongly, have
substantial weight in the path integral. The former give rise to the Hubbard
bands, while the latter ones are responsible for the low--frequency Kondo
resonance. Two features in these spectra are surprpizing. First, the narrow
central peak before the Mott transition, resulting from quasiparticle states
formed in the background of the coherent Kondo tunneling of the local spin
fluctuations. Second, atomic physics leaves a signature on the one--particle
spectral properties in the form of well--formed Hubbard bands at higher
frequency, in the strongly--correlated metallic state.

As the transition at zero temperature is approached, there is a substantial
transfer of spectral weight from the low--lying quasiparticles to the
Hubbard bands. The Mott transition at zero temperature takes place at a
critical value of $U$, denoted by $U_{c2}$ where the integrated spectral
weight at low frequency vanishes, as shown in figure \ref{spectralu}. This
results in a Mott transition point where the quasiparticle mass diverges,
but a discontinuous gap opens in the quasiparticle spectra\cite{zrk} . These
results are in agreement with the early work of Fujimori {\it et al.} \cite
{fujimori}, who arrived at essentially the same picture on the basis of
experiment. It is worth remarking that the spectral function in the strongly
correlated metallic region is better regarded as composed of three
components: Hubbard bands centered at $U/2$, a low--energy quasiparticle
peak with a height of order unity and a total intensity proportional to $({%
U_{c2}}-U)$ distributed over an energy range $U_{c2}-U$, and an incoherent
background connecting the high energy to the low--energy region. The last
feature ensures that there is no real gap between the quasiparticle features
and the Hubbard bands, as long as one stays in the metallic regime.

\subsubsection{Anomalous resistivities}

\begin{figure}[tbp]
\center{\epsfig{file=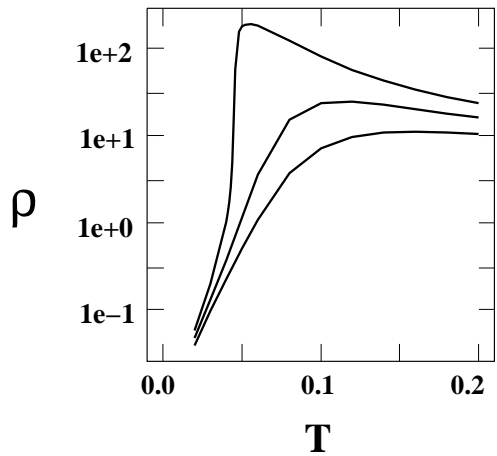,height=60mm,angle=0}}
\caption{$\protect\rho _{dc}(T)$ around the coherence incoherence crossover
near the finite temperature Mott endpoint. $U/D=2.1,2.3,2.5$ (bottom to
top), obtained with the IPT method from ref \protect\onlinecite{rozenberg}.}
\label{resist}
\end{figure}

Figure \ref{resist} describes the anomalous resistivities near these
crossover regions. Notice the anomalously large metallic resistivity which
is typical of many oxides \cite{kivelson}. While the curves in this figure
far exceed the Ioffe--Regel limit (using estimates of $k_{f}$ from T=0
calculations) there is no violation of any physical principle. At low
temperatures, a ${\bf k}$ space based Fermi--liquid--theory description
works but in this regime the resistivity is low (below the Ioffe--Regel
limit). Above certain temperature the resistivity exceeds the Ioffe--Regel
limit but then quasiparticle description becomes inadequate. There is no
breakdown or singularities in our formalism, the spectral functions remain
smooth, (above the Mott--transition endpoint), only the physical picture
changes. At high temperatures we have an incoherent regime to which the
Ioffe--Regel criteria does not apply, because there are no long--lived
excitations with well--defined crystal momentum in the spectra. The electron
is strongly scattered off orbitals and spin fluctuations, and is better
described in real space. In this regime, there is no simple description in
terms of \ ${\bf k}$ space elementary excitations, but one can construct a
simple description and perform quantitative calculations if one adopts the
spectral function as a basic object in terms of which one formulates the
theory.

\begin{figure}[tbp]
\center{\epsfig{file=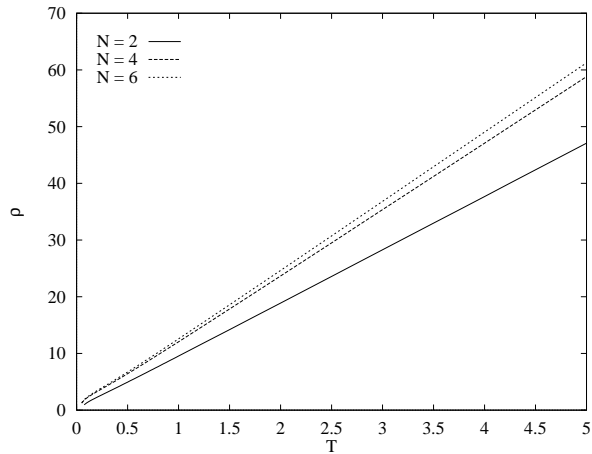,height=60mm,angle=0}}
\caption{$\protect\rho _{dc}(T)$ in units of $z \hbar a/{e^2}$, vs T (in
units of D) for different values of orbital degeneracy N for a fixed doping $%
\protect\delta=.1$ obtained with the NCA method which is valid at high
temperatures, from ref \protect\onlinecite{palsson}.}
\label{resist3}
\end{figure}

\begin{figure}[tbp]
\center{\epsfig{file=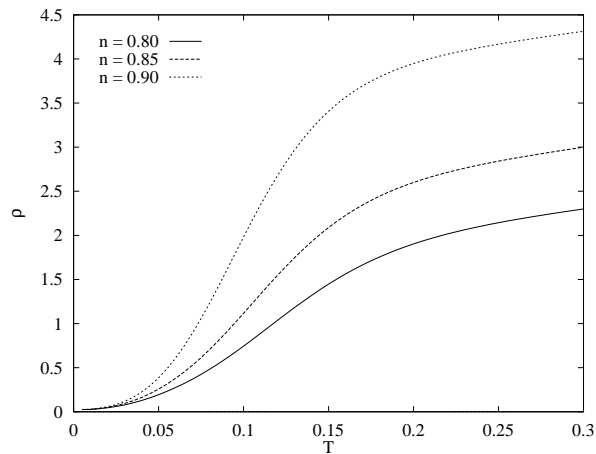,height=60mm,angle=0}}
\caption{$\protect\rho _{dc}(T)$ vs T in units of $z \hbar a/{e^2}$, vs T
(in units of D) obtained with the IPT method, for different dopings at ${%
\frac{U }{D}} =2.8$ from ref \protect\onlinecite{palsson}.}
\label{resist2}
\end{figure}

Only the anomalously large magnitude of the resistivity (which follows from
the Green function which has branch cuts rather than well--defined poles),
is universal as can be seen by comparison of the detailed temperature
dependence at half filling (as in fig \ref{resist}) and away from half
filling as in figures \ref{resist3} and \ref{resist2}. The temperature
dependence of the transport in the high--temperature incoherent regime
depends on whether the system is at integer filling or doped, as can be
shown numerically \cite{jarrell} and analytically \cite{palsson} in the
example of the doped Mott insulator. The low--temperature and the
high--temperature anomalously large resistivities also occur in strongly
coupled electron--phonon systems, as discussed earlier\cite{millis}.

\subsubsection{Anomalous Transfer of Spectral Weight}

Another manifestation of the same physics is the anomalous transfer of
spectral weight which is observed in the one--electron and in the optical
spectra of correlated systems as parameters such as doping or pressure are
varied. This surprising aspect of strong--correlation physics was noted and
emphasized by many authors\cite{sawatzki}. Transfer of spectral weight can
also take place as a function of temperature. For example the ''kinetic
energy'' which appears in the low--energy optical sum rule can have sizeable
temperature dependence, an effect that was discovered experimentally by \cite
{schlessinger} and\ explained theoretically by DMFT\ calculations \cite
{marcelooptics}

\begin{figure}[htb]
\centerline{\epsfig{file=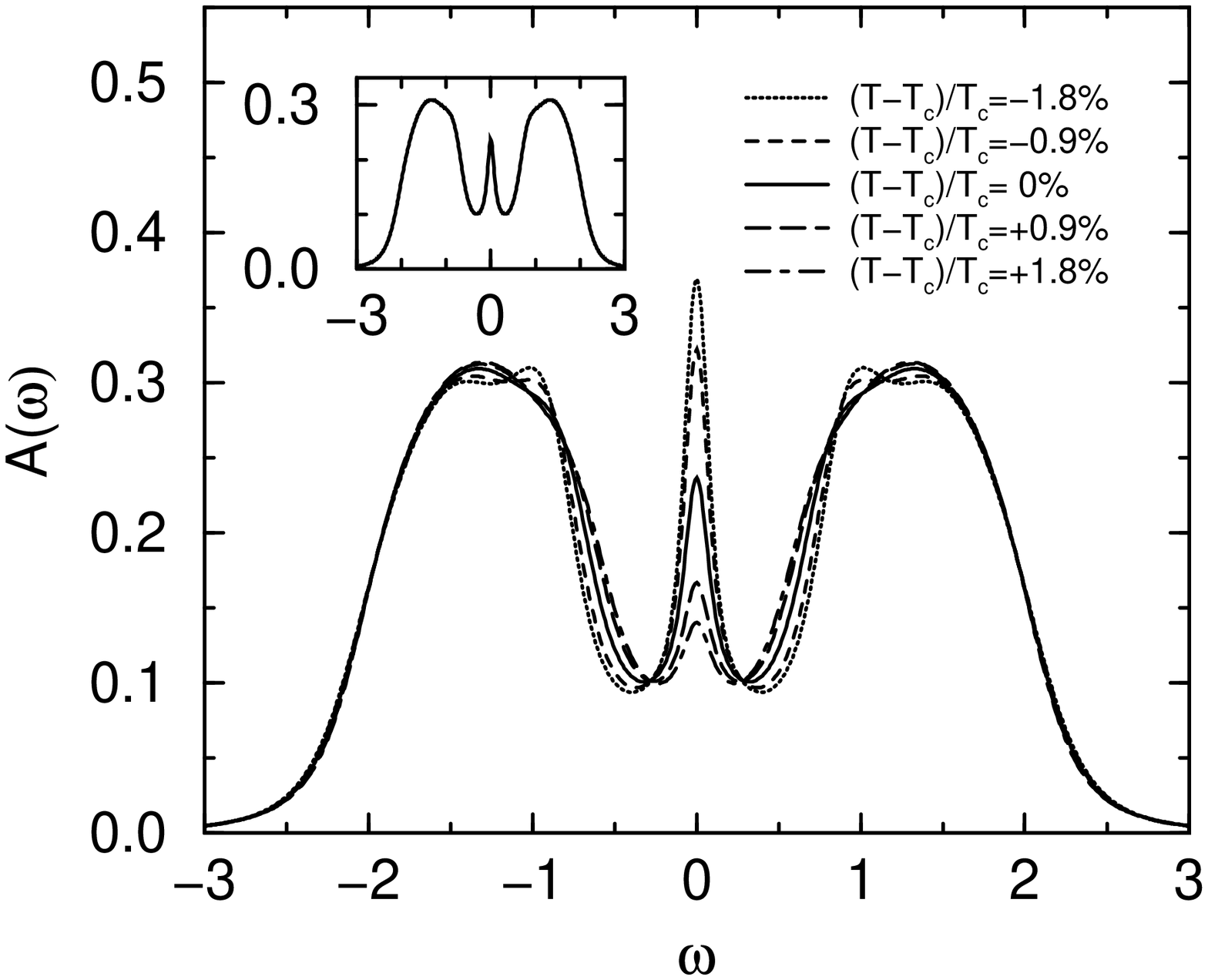,width=3.5in}}
\caption{Evolution of the spectral function as function of temperature
(bottom to top), near the finite temperature from Ref. \protect\onlinecite{lange}.
The inset is the spectral function at the second order Mott endpoint. }
\label{spectralt}
\end{figure}

Once more, thinking about this problem in terms of well--defined
quasiparticles is not useful. It is more fruitful to formulate the problem \
in terms of spectral functions describing on the same footing coherent and
incoherent excitations. The relative weights of these components in the
spectra evolves smoothly with temperature and leads to sizeable variations
in the integrated optical intensity. The evolution of the spectral function
near the temperature driven Mott transition is shown in figure \ref
{spectralt}.

The physics of strongly correlated materials in a wide range of parameters
cannot be described in terms of quasiparticle excitations. Recent advances
in the theory of the Mott transition highlight the fundamental role of the
spectral function. The functionals discussed in sections VII, VIII, and XII
carry over this ideology to realistic electronic--structure calculations.

\section{Density Functional Theory and LDA}

\label{lda}Density functional theory in the Kohn--Sham formulation is one of
the basic tools for studying weakly--interacting \ electronic systems as is
widely used by the electronic--structure community. We review it briefly
here using an effective action point of view in order to highlight the
similarities and differences with the DMFT methods which will be presented
in the same language. The approach in this context was introduced by Fukuda 
\cite{fukuda,aliev,argaman} and it amounts to a simple Legendre
transformation. One introduces the generating functional of the connected
Green functions in the presence of an arbitrary potential.

Consider the partition function $Z$ (or equivalently the free energy $W$) of
a system of electrons moving in a crystal potential $V_{ext}(x)$ and
interacting via Coulomb interactions $V$, in the presence of an external
source $J$ coupled to the electron density 

\begin{eqnarray}
Z=\exp [-W[J]] &=&\int D[\psi \psi ^{\dagger }]\exp [-S
- \int dx J(x) {{\psi_\sigma}^{\dagger }} (x)\psi_\sigma (x)]
\end{eqnarray}

\begin{eqnarray}
S &=&\int dx\sum_{\alpha }\psi _{\sigma }^{+}(x)[\partial _{\tau }-{\frac{{%
\bigtriangledown ^{2}}}{{2m}}}+V_{ext}(x)]\psi _{\sigma }(x) \nonumber \\
&+&{\frac{{e^{2}}}{2}}\sum_{\alpha \beta }\int dxdx^{\prime }\psi _{\sigma
}^{+}(x)\psi _{\sigma ^{\prime }}^{+}(x^{\prime })U_{c}(x-x^{\prime })\psi
_{\sigma ^{\prime }}(x^{\prime })\psi _{\sigma }(x)  \label{iham}
\end{eqnarray}
In (\ref{iham}), $x=({\bf x},\tau )$ denote space--imaginary time
coordinates, $V_{ext}$ is the crystal  
potential created by the ions and  $%
U_{c}(x-x^{\prime })=|{\bf x}-{\bf x^{\prime }}|^{-1}\delta (\tau -\tau
^{\prime })$ is the Coulomb interaction.

The density functional 
$\ \Gamma (\rho )$  is defined by

\begin{equation}
\Gamma \lbrack \rho ]=W[J]-\int J(x)\rho ({\bf x})  \label{bk1}
\end{equation}

where  one eliminates the source  in favor of the density.
The minimum  of the density functional gives the true density
and the total energy of the solid.

To construct approximations to the functional $\Gamma $ it is very useful to
introduce the Kohn--Sham potential, $V_{KS},$ which is defined as the
potential such that added to the non--interacting kinetic energy produces
the given density in a reference system of non--interacting particles . i.e.

\begin{equation}
\rho ({\bf r})=T\sum_{\sigma }\sum_{i\omega _{n}}\langle \sigma {\bf r}%
\left| (i\omega _{n}+\nabla ^{2}/2-V_{KS})^{-1}\right| \sigma {\bf r}\rangle
e^{i\omega_n 0^{+}}  \label{rhov}
\end{equation}

The exact functional can be viewed as a functional of two variables 
\begin{eqnarray}
\Gamma (\rho ,V_{KS}) &=&-T\sum_{i\omega _{n}}tr\log [i\omega _{n}+\nabla
^{2}/2-V_{KS}]-\int V_{KS}({\bf r})\rho ({\bf r})d{\bf r+}  \label{grhov} \\
&&\frac{1}{2}\int \frac{\rho ({\bf r})\rho ({\bf r}^{\prime })}{|{\bf r}-%
{\bf r}^{\prime }|}d{\bf r}d{\bf r}^{\prime }+\int V_{ext}({\bf r})\rho (%
{\bf r})d{\bf r+}E_{xc}[\rho ]  \nonumber
\end{eqnarray}
$\Gamma (\rho )$ is obtained by substituting $V_{KS}(\rho )$ obtained by
solving Eq. (\ref{rhov}) (which makes (\ref{grhov}) stationary) onto $\Gamma
(\rho ,V_{KS}).$ $E_{xc}[\rho ]$ is the exchange--correlation energy which
is a functional of the density and not of the external potential.

Extremizing (\ref{grhov}) with respect to $\rho $ gives 
\begin{equation}
V_{KS}({\bf r})[\rho ]=\int \frac{\rho ({\bf r}^{\prime })d{\bf r}^{\prime }%
}{|{\bf r}-{\bf r}^{\prime }|}+V_{xc}({\bf r})[\rho ]+V_{ext}({\bf r})[\rho ]
\label{vks}
\end{equation}
where $V_{xc}({\bf r})$ is the exchange-correlation potential obtained as 
\begin{equation}
\frac{\delta E_{xc}}{\delta \rho ({\bf r})}\equiv V_{xc}({\bf r})
\label{vxc}
\end{equation}
Since $E_{xc}[\rho ]$ is not known explicitly approximations are needed. The
LDA assumes 
\begin{equation}
E_{xc}[\rho ]=\int \epsilon _{xc}[\rho ({\bf r})]\rho ({\bf r})d{\bf r}
\label{exclda}
\end{equation}
with $\epsilon _{xc}[\rho ({\bf r})]$ being the energy density of the
uniform electron gas, a function which is easily parametrized.

For simplicity we restrict ourselves to zero--temperature and interpret the
Fermi functions as the Heaviside step functions. Eq. (\ref{rhov}) can be
rewritten as the eigenvalue problem 
\begin{equation}
\lbrack -\nabla ^{2}/2+V_{KS}({\bf r})]\psi _{{\bf k}j}({\bf r})=\epsilon _{%
{\bf k}j}\psi _{{\bf k}j}({\bf r})  \label{ks}
\end{equation}
\begin{equation}
\rho ({\bf r})=\sum_{{\bf k}j}f(\epsilon _{{\bf k}j})\psi _{{\bf k}j}^{\ast
}({\bf r})\psi _{{\bf k}j}({\bf r})  \label{rho}
\end{equation}
and $V_{KS}$ is given as an explicit function of the density. In practice
one frequently uses the analytical formulae\cite{PerdewWang}. The idea here
is to fit a functional form to a Quantum--Monte--Carlo (QMC)\ calculation,
and we will return to this idea when we discuss the IPT.

Then, the total energy of the crystal is given as 
\begin{equation}
E_{tot}=\sum_{{\bf k}j}f(\epsilon _{{\bf k}j})\epsilon _{{\bf k}j}+\frac{1}{2%
}\int \frac{\rho ({\bf r})\rho ({\bf r}^{\prime })}{|{\bf r}-{\bf r}^{\prime
}|}d{\bf r}d{\bf r}^{\prime }+\int V_{ext}({\bf r})\rho ({\bf r})d{\bf r+}%
\int \epsilon _{xc}[\rho ({\bf r})]\rho ({\bf r})d{\bf r}+E_{dc}
\label{EtotDFT}
\end{equation}
where 
\begin{equation}
E_{dc}=-\int V_{KS}({\bf r})\rho ({\bf r})d{\bf r}  \label{EdcDFT}
\end{equation}
simply subtract the interaction energy from the Kohn--Sham eigenvalues which
is explicitly included in the Hartree and exchange--correlation term to
avoid double counting.

Density functional is an exact approach as long as the invertibility
condition expressing the potential in terms of the density is satisfied.
However in strongly correlated situations, the total energy is not very
sensitive to the potential since the electrons are localized due to the
interactions themselves, and the lack of sensitivity of the functional to
the density, hampered the design of good approximations to the exact
functional in this regime. Furthermore, at the Mott transition the
invertibility condition may not be satisfied.

The effective action approach exhibits that the functional $\Gamma $ is a
Legendre transform of the exact generating functional $W[J]$ and introduces
Kohn--Sham field in a perturbative expansion of the Legendre transform in $%
e^{2}$ at fixed density.

The Kohn--Sham parametrization of the density in terms of $V_{KS}$ is
extremely useful because it expresses uniquely $\rho ({\bf r})$ in terms of
the Kohn--Sham orbitals $\psi _{{\bf k}j}({\bf r}).$ Truncations of DFT are
necessary for practical implementations. There are two different
philosophies in truncating Eqs.(\ref{vks})--(\ref{EtotDFT}), one is via the
introduction of pseudopotentials which we will not describe here.
All--electron methods simply introduce a finite basis set $\chi _{\alpha }^{%
{\bf k}}({\bf r})$ and expand 
\begin{equation}
\psi _{{\bf k}j}({\bf r})=\sum_{\alpha }\chi _{\alpha }^{{\bf k}}({\bf r}%
)A_{\alpha }^{{\bf k}j}  \label{psikjlmto}
\end{equation}
keeping a finite set of $\alpha .$ Notice that this truncation immediately
restricts the active part of the multiplicative operator associated with the
Kohn--Sham potential to have a form \ 
\begin{equation}
\hat{V}=\sum_{{\bf k}}|\chi _{\alpha }^{{\bf k}}\rangle V_{\alpha \beta
}\langle \chi _{\alpha }^{{\bf k}}|  \label{ksop}
\end{equation}

Of course, one can add to this contributions from the set which is
orthogonal to the minimal basis set $|\chi _{\alpha }^{{\bf k}}\rangle $
without changing the truncated density. The requirement of locality
presumably determines the Kohn--Sham potential and the component of (\ref
{ksop}) outside the space of$\chi _{\alpha }^{{\bf k}}({\bf r})$ uniquely.

Linear muffin--tin orbitals (LMTO's) \cite{LMTO} are an optimal minimal
basis set. For a known Kohn--Sham potential this construction can be done
once and for all. However, since $V_{KS}$ depends on the density, the basis $%
|\chi _{\alpha }^{{\bf k}}\rangle $ is adapted iteratively to the
self--consistent solution. The same observation applies to our new
implementation of the LDA+DMFT approach.

In principle, density functional theory is an exact theory as long as the
invertibility condition expressing the potential in terms of the density is
satisfied. In practice, the usefulness of this approach is due to the
existence of very successful approximations to the exchange--correlation
potential, by the LDA or the GGA. In principle the Kohn--Sham orbitals is a
technical device for generating the total energy, however in practice they
are used as a first step in perturbative calculations of the one--electron
Green function, as in the GW method. The LDA method is very successful in
many materials for which the standard model of solids works. However, in
correlated electron system this is not always the case. Our view, is that
this situation cannot be remedied by using more complicated exchange and
correlation functionals in density functional theory. As discussed in the
previous section the spectra of strongly correlated electron systems have
both Hubbard bands (which have no analog in one--electron theory) and
correlated quasiparticle bands in the one--electron spectra. DMFT is the
simplest approach which is based on this simple physical picture.

The extension to spin--density functional, in its non--collinear version is
straightforward. The functional $\Gamma $ should now be considered as a
functional of both the charge density $\rho (r)$ and magnetization density $%
{\bf m}({\bf r})$ The latter is a vector quantity. It can be slightly
non--collinear due to spin--orbit coupling effects. The functional of the
total energy, Eq.(\ref{EtotDFT}) is rewritten by taking into account the
fact that the Kohn--Sham field now consists of both the potential $V_{KS}(%
{\bf r})$ and magnetic field ${\bf B}_{KS}({\bf r})$

\begin{eqnarray}
E_{tot} &=&\sum_{{\bf k}j}f(\epsilon _{{\bf k}j})\epsilon _{{\bf k}j}+\frac{1%
}{2}\int \frac{\rho ({\bf r})\rho ({\bf r}^{\prime })}{|{\bf r}-{\bf r}%
^{\prime }|}d{\bf r}d{\bf r}^{\prime }+\int V_{ext}({\bf r})\rho ({\bf r})d%
{\bf r+\int B_{ext}(r)m(r)dr}  \label{EtotSDFT} \\
&&{\bf +}\int \epsilon _{xc}[\rho ({\bf r}),|{\bf m}({\bf r})|]\times \rho (%
{\bf r})d{\bf r+}\int f_{xc}[\rho ({\bf r}),|{\bf m}({\bf r})|]\times |{\bf m%
}({\bf r})|d{\bf r}+E_{dc}  \nonumber
\end{eqnarray}
where we have used the local spin density approximation (LSDA) expression
for exchange--correlation energy by assuming that the properties of electron
gas depend on the absolute value of the magnetization: $|{\bf m}({\bf r})|$.

The double--counting term is 
\begin{equation}
E_{dc}=-\int V_{KS}({\bf r})\rho ({\bf r})d{\bf r-}\int {\bf B}_{KS}({\bf r})%
{\bf m}({\bf r})d{\bf r}  \label{EdcSDFT}
\end{equation}
with 
\begin{eqnarray}
{\bf B}_{KS}({\bf r}) &=&{\bf B}_{ext}({\bf r})+\frac{\delta E_{xc}}{\delta 
{\bf m}({\bf r})}  \label{bks} \\
{\bf m}({\bf r}) &=&\sum f_{{\bf k}j}\langle \vec{\psi}_{{\bf k}j}|{\bf \hat{%
s}}|\vec{\psi}_{{\bf k}j}\rangle _{spin}  \label{mks}
\end{eqnarray}
(we average over spin degrees of freedom only), and spinor wave functions $%
\vec{\psi}_{{\bf k}j}$ satisfy to the Pauli--like Kohn--Sham matrix equation:

\[
\lbrack (-\nabla ^{2}+V_{KS}({\bf r}))\hat{I}+{\bf B}_{KS}({\bf r}){\bf \hat{%
s}}+\xi {\bf \hat{l}\hat{s}}]\vec{\psi}_{{\bf k}j}=\epsilon _{{\bf k}j}\vec{%
\psi}_{{\bf k}j} 
\]
where $\hat{I}$ is the unity 2*2 matrix; ${\bf s}$ is a spin operator which
is just the Pauli matrix divided by two; $\xi $ determines the strength of
spin--orbit coupling and in practice is determined\cite{Harmon} by radial
derivative of the $l=0$ component of the Kohn--Sham potential inside an
atomic sphere: 
\begin{equation}
\xi (r)=\frac{2}{c^{2}}\frac{dV_{KS}(r)}{dr}  \label{sor}
\end{equation}

\section{LDA+U\ Method}

We now turn to a description of the LDA+U method\cite{LDA+U}. We will deal
with the spin unrestricted formulation and for illustration purposes assume
that relativistic effects are small. This allows us to choose a quantization
axis along some direction, say $z$, since the total energy is now invariant
with respect to its orientation. Instead of considering $\rho ({\bf r})$ and
full vector ${\bf m}(r)$ we can deal with $\rho ({\bf r})$ and only $m_{z}(%
{\bf r})$ or, equivalently with spin--up and --down densities $\rho ^{\sigma
}({\bf r})=[\rho ({\bf r})+\sigma m_{z}({\bf r})]/2$ , $\sigma =\pm 1.$

The approach requires an introduction of a set of localized orbitals $\phi
_{a}({\bf r}-{\bf R})$ which are used to build an ``occupancy spin density
matrix'' 
\begin{equation}
n_{ab}^{\sigma }=\sum_{{\bf k}j}f(\epsilon _{{\bf k}j\sigma })\int \psi _{%
{\bf k}j\sigma }^{\ast }({\bf r})\phi _{a}({\bf r})d{\bf r}\int \psi _{{\bf k%
}j\sigma }({\bf r}^{\prime })\phi _{b}^{\ast }({\bf r}^{\prime })d{\bf r}%
^{\prime }  \label{dmat}
\end{equation}
This intuitively represents the ''correlated part of the electron density''
as long as we associate our projectors $\phi _{a}$ with correlated electrons
. The total energy now is represented as a functional of the spin densities $%
\rho ^{\sigma }({\bf r})$ and of $n_{ab}^{\sigma }.$ In complete analogy
with Eq. (\ref{grhov}), one introduces a Lagrange multipliers matrix $%
\lambda _{ab}^{\sigma }$ such to enforce (\ref{dmat}) and to expand the
LDA+U functional as 
\begin{eqnarray}
&&\Gamma _{LDA+U}[n_{ab}^{\sigma },\lambda _{ab}^{\sigma },V_{KS}^{\sigma
},\rho ^{\sigma }]=-T\sum_{i\omega _{n}}tr\log [i\omega _{n}+\nabla
^{2}/2-V_{KS}^{\sigma }-\sum_{ab}\lambda _{ab}^{\sigma }\phi _{a}({\bf r}%
)\phi _{b}^{\ast }({\bf r}^{\prime })]-  \label{glda+u} \\
&&\sum_{\sigma }\int V_{KS}^{\sigma }({\bf r})\rho ^{\sigma }({\bf r})d{\bf r%
}-\sum_{\sigma }\sum_{ab}\lambda _{ab}^{\sigma }n_{ab}^{\sigma }+\frac{1}{2}%
\int \frac{\rho ({\bf r})\rho ({\bf r}^{\prime })}{|{\bf r}-{\bf r}^{\prime
}|}d{\bf r}d{\bf r}^{\prime }+E_{xc}^{LDA}[\rho ^{\sigma }]+ \\
&&E^{Model}[n^{\sigma }]-E_{dc}^{Model}[n^{\sigma }]  \nonumber
\end{eqnarray}
where we have added a contribution from the Coulomb energy in the shell of
correlated electrons 
\begin{equation}
E^{Model}[n^{\sigma }]=\frac{1}{2}\sum_{\sigma
}\sum_{abcd}U_{abcd}n_{ab}^{\sigma }n_{cd}^{-\sigma }+\frac{1}{2}%
\sum_{\sigma }\sum_{abcd}(U_{abcd}-J_{abcd})n_{ab}^{\sigma }n_{cd}^{\sigma }
\label{emodelldau}
\end{equation}
Since, partially this energy is already taken into account in LDA, we have
to subtract a double-counting part denoted by $E_{dc}^{Model}[n^{\sigma }].$

Minimization of this functional gives rise to 
\begin{equation}
\rho ^{\sigma }({\bf r})=T\sum_{i\omega _{n}}\left\langle {\bf r}\left|
[i\omega _{n}+\nabla ^{2}/2-V_{KS}^{\sigma }-\sum_{ab}\lambda _{ab}^{\sigma
}\phi _{a}({\bf r})\phi _{b}^{\ast }({\bf r}^{\prime })]^{-1}\right| {\bf r}%
\right\rangle e^{i\omega _{n}0^{+}}=\sum_{{\bf k}j}f(\epsilon _{{\bf k}%
j\sigma })|\psi _{{\bf k}j\sigma }({\bf r})|^{2}  \label{rholda+u}
\end{equation}
with 
\begin{equation}
\lbrack -\nabla ^{2}/2+V_{KS}^{\sigma }+\sum_{ab}\lambda _{ab}^{\sigma }\phi
_{a}({\bf r})\phi _{b}^{\ast }({\bf r}^{\prime })]\psi _{{\bf k}j\sigma
}=\epsilon _{{\bf k}j\sigma }\psi _{{\bf k}j\sigma }  \label{kslda+u}
\end{equation}
where $V_{KS}^{\sigma }$ is given by equation similar to Eq.(\ref{vks}) and 
\begin{equation}
\lambda _{ab}^{\sigma }=\sum_{cd}U_{abcd}n_{cd}^{-\sigma
}+\sum_{cd}(U_{abcd}-J_{abcd})n_{cd}^{\sigma }-\frac{dE_{dc}^{Model}[n^{%
\sigma }]}{dn_{ab}^{\sigma }}  \label{lambdalda+u}
\end{equation}
Several remarks are in order.

(i) The LDA+U functional and the LDA+U equations are defined once a set of
projectors and a matrix of interactions $U_{abcd}$ is prescribed. Formally,
the matrices $\hat{U}$ and $\hat{J}$ have the following definitions:

\begin{eqnarray}
U_{abcd} &=&\langle ac|\frac{1}{r}|bd\rangle =\int_{a}^{\ast }{\phi^*}_{a}(%
{\bf r})\phi _{c}^{\ast }({\bf r}^{\prime })v_{C}({\bf r}-{\bf r}^{\prime
})\phi _{b}({\bf r})\phi _{d}({\bf r}^{\prime })d{\bf r}d{\bf r}^{\prime }
\label{C2} \\
J_{abcd} &=&\langle ac|\frac{1}{r}|db\rangle =\int_{a}^{\ast }{{\phi_a}^*}(%
{\bf r})\phi _{c}^{\ast }({\bf r}^{\prime })v_{C}({\bf r}-{\bf r}^{\prime
})\phi _{d}({\bf r})\phi _{b}({\bf r}^{\prime })d{\bf r}d{\bf r}^{\prime }
\label{C3}
\end{eqnarray}
where the Coulomb interaction $v_{C}({\bf r}-{\bf r}^{\prime })$ has to take
into account the effects of screening by conduction electrons. In practice,
one can express these matrices via a set of Slater integrals which, for
example, for d--electrons are parametrized by three constants $F^{(0)},F^{(2)},$
and $F^{(4)}.$

An important question is to discuss the double counting term $%
E_{dc}^{Model}[n^{\sigma }]$ which has been occurred when we added
additional Coulomb interaction to the functional. It was proposed\cite
{Liecht} that the form for $E_{dc}^{Model}[n^{\sigma }]$ is

\begin{equation}
E_{dc}^{Model}=\frac{1}{2}\bar{U}\bar{n}(\bar{n}-1)-\frac{1}{2}\bar{J}[\bar{n%
}^{\uparrow }(\bar{n}^{\uparrow }-1)+\bar{n}^{\downarrow }(\bar{n}%
^{\downarrow }-1)].  \label{C6}
\end{equation}
where 
\begin{eqnarray}
\bar{U} &=&\frac{1}{(2l+1)^{2}}\sum_{ab}\langle ab|\frac{1}{r}|ab\rangle
\label{C7} \\
\bar{J} &=&\bar{U}-\frac{1}{2l(2l+1)}\sum_{ab}(\langle ab|\frac{1}{r}%
|ab\rangle -\langle ab|\frac{1}{r}|ba\rangle )  \label{C8}
\end{eqnarray}
and where $\bar{n}^{\sigma }=\sum_{a}n_{aa}^{\sigma },$ and $\bar{n}=\bar{n}%
^{\uparrow }+\bar{n}^{\downarrow }.$ This generates the correction to the
potential in the form: 
\begin{equation}
\lambda _{ab}^{\sigma }=\sum_{cd}U_{abcd}n_{cd}^{-\sigma
}+\sum_{cd}(U_{abcd}-J_{abcd})n_{cd}^{\sigma }-\bar{U}(\bar{n}-\frac{1}{2}%
)+J(\bar{n}^{\sigma }-\frac{1}{2})  \label{potldau}
\end{equation}

The question now arises as to whether the double--counting correction should
subtract self--interaction effects or not. In principle, if the density
functional contains this spurious term, the same should be taken into
account in the double--counting expression. Judging by the experience that
the LDA total energy is essentially free of self--interaction (the total
energy of the hydrogen atom is, for example, very close to --1 Ry, while the
Kohn--Sham eigenvalue is only about --0.5 Ry), the construction $%
E_{dc}^{Model}$ is made so as to be free of the self--interaction. However
this statement cannot be considered seriously in general, and alternative
forms of the double counting term includiing the effects of
self--interaction have been used\cite{doublec}

(iii) If one uses the Eqs.(\ref{vks}), (\ref{lambdalda+u}) and (\ref{dmat})
to eliminate $V_{KS}^{\sigma }$, $n_{ab}^{\sigma }$, and $\lambda
_{ab}^{\sigma }$ as functions of $\rho ^{\sigma }$, and backsubstitutes that
into (\ref{glda+u}), one obtains a functional of the density alone.

(iv) The literature is ambiguous on whether $\Gamma _{LDA+U}$ could perhaps
be viewed as an approximation to a functional of \ a projected density
matrix , or, as indicated in (iii), simply as a different density functional
where in addition to dividing the density into spin up and spin down as in
LSDA, one introduce a correlated component (i.e. what is parametrized by $%
n_{ab}$) of the density and an uncorrelated one.

(v) A different point of view is to introduce a "correlated part of the
one--particle density matrix" 
\begin{equation}
\int \int \phi _{a}({\bf r})\langle \psi ^{+}({\bf r})\psi ({\bf r}^{\prime
})\rangle \phi _{b}({\bf r}^{\prime })d{\bf r}d{\bf r}^{\prime }
\label{dmat2}
\end{equation}
and to consider a functional of this quantity and of the total density by
effective action methods and view Eq. (\ref{glda+u}) as an approximation to
this exact functional. However, the interpretation of 
\begin{equation}
T\sum_{i\omega _{n}}\left\langle {\bf r}^{\prime }\left| [i\omega
_{n}+\nabla ^{2}/2-V_{KS}^{\sigma }-\sum_{ab}\lambda _{ab}^{\sigma }\phi
_{a}({\bf r})\phi _{b}^{\ast }({\bf r}^{\prime })]^{-1}\right| {\bf r}%
\right\rangle e^{i\omega _{n}0^{+}}  \label{dmat3}
\end{equation}
as a density matrix is not consistent with existence of interactions \ This
is because Eq. (\ref{dmat2}) describes a density matrix which has
eigenvalues which are ones and zeros, and this characterizes a
non--interacting density matrix. The density matrix of an interacting system
has eigenvalues which are less than one. Therefore, Eq.(\ref{dmat}) can not
represent an interacting density matrix. Because of these difficulties we
have suggested to interpret the LDA+U\ method as a static limit of the more
powerful DMFT\ method which we describe later. The static limit of DMFT is
going to be most accurate as more symmetries (spin, orbital) are broken.
Removing local degeneracies by spontaneous symmetry breaking is the simplest
way of minimizing the energy (reducing the correlations) \cite{Brandow}.

(vi) It was argued\cite{ferdi} that the Green function

\begin{equation}
\left\langle {\bf r}^{\prime }\left| [i\omega _{n}+\nabla
^{2}/2-V_{KS}^{\sigma }-\sum_{ab}\lambda _{ab}\phi _{a}({\bf r})\phi
_{b}^{\ast }({\bf r}^{\prime })]^{-1}\right| {\bf r}\right\rangle
\label{gfunction}
\end{equation}
can be viewed as a limiting case of the GW approximation\cite{hedin} but
this is again not clear since an interacting Green's function has poles with
residues less than one, and this is not the case in expression (\ref
{gfunction}) except for the uncorrelated situation when Hartree Fock theory
becomes exact.

(vii) Relativistic effects which are important for such applications such as
magnetic anisotropy calculations, can be considered. We have described the
extended DFT in the previous section, and here we only discuss the LDA+U
corrections. If spin--orbit coupling is taken into account, the occupancy
matrix becomes non--diagonal with respect to spin index: 
\begin{equation}
n_{ab}^{\sigma \sigma ^{\prime }}=\sum_{{\bf k}j}f(\epsilon
_{{\bf k}j})\int \psi _{{\bf k}j}^{\ast \sigma }({\bf r})\phi _{a}({\bf r})d%
{\bf r}\int \psi _{{\bf k}j}^{\sigma ^{\prime }}({\bf r}^{\prime })\phi
_{b}^{\ast }({\bf r}^{\prime })d{\bf r}^{\prime }  \label{dmat4}
\end{equation}
The correction to the functional has the form similar to Eq(\ref{emodelldau}%
) and it is given by 
\begin{equation}
E^{Model}[n^{\sigma \sigma ^{\prime }}]=\frac{1}{2}\sum_{abcd\sigma
}U_{abcd}n_{ab}^{\sigma \sigma }n_{cd}^{-\sigma -\sigma }+\frac{1}{2}%
\sum_{abcd\sigma }(U_{abcd}-J_{abcd})n_{ab}^{\sigma \sigma }n_{cd}^{\sigma
\sigma }-\frac{1}{2}\sum_{abcd\sigma }J_{abcd}n_{ab}^{\sigma -\sigma
}n_{cd}^{-\sigma \sigma }  \label{Emodel}
\end{equation}
which can be worked out by considering a Hartree--Fock average of the
original expression for the Coulomb interaction 
\begin{equation}
\frac{1}{2}\sum_{\sigma \sigma ^{\prime }}\sum_{abcd}\langle a\sigma b\sigma
^{\prime }|\frac{e^{2}}{r}|c\sigma d\sigma ^{\prime }\rangle c_{a\sigma
}^{+}c_{b\sigma ^{\prime }}^{+}c_{d\sigma ^{\prime }}c_{c\sigma }
\label{coulomb}
\end{equation}
The correction to the potential takes the same form as Eq.(\ref{potldau})
when $\sigma \equiv \sigma ^{\prime }$, i.e., 
\begin{equation}
\lambda _{ab}^{\sigma \sigma }=\sum_{cd}U_{abcd}n_{cd}^{-\sigma -\sigma
}+\sum_{cd}(U_{abcd}-J_{abcd})n_{cd}^{\sigma \sigma }-\bar{U}(\bar{n}-\frac{1%
}{2})+J(\bar{n}^{\sigma \sigma }-\frac{1}{2})  \label{lambdasigma}
\end{equation}
and for the off--diagonal elements it is given by 
\begin{equation}
\lambda _{ab}^{\sigma -\sigma }=-\sum_{cd}J_{abcd}n_{cd}^{-\sigma \sigma }
\label{lambdaoff}
\end{equation}
To make it more physically transparent we can introduce magnetic moments at
the given shell by 
\begin{equation}
m_{ab}^{\mu }=\sum_{\sigma \sigma ^{\prime }}s_{\sigma \sigma ^{\prime
}}^{\mu }n_{ab}^{\sigma \sigma ^{\prime }}  \label{magmom}
\end{equation}
where $\mu $ runs over $x,y,z$ for Cartesian coordinates, or over,$-1,0,+1$ (%
$z,\pm $) for spherical coordinates. Relativistic correction to the LDA+U
energy can be written in physically transparent form 
\begin{equation}
\frac{1}{2}\sum_{abcd\sigma }J_{abcd}n_{ab}^{\sigma -\sigma }n_{cd}^{-\sigma
\sigma }\equiv \frac{1}{2}\sum_{abcd}m_{ab}^{(+)}J_{abcd}m_{cd}^{(-)}+\frac{1%
}{2}\sum_{abcd}m_{ab}^{(-)}J_{abcd}m_{cd}^{(+)}  \label{phystrans}
\end{equation}
In principle one can assume further generalization of exchange matrix $%
J_{abcd}$ to be anisotropic, i.e depending on $\mu \mu ^{\prime }$: $%
J_{abcd}^{\mu \mu ^{\prime }}.$

To summarize, since the density uniquely defines the Kohn--Sham orbitals,
and they in turn, determine the occupancy matrix of the correlated orbitals
[once a choice of correlated orbital in Eq. (\ref{dmat4})] is made, we still
have a functional of the density alone. However it is useful to proceed by
analogy with Eq. (\ref{glda+u}), and think of the LDA + U functional as a
functional of $\rho ^{\sigma }$, $n^{\sigma }$ $V_{KS}^{\sigma }$ and $%
\lambda ^{\sigma }$, whose minimum gives better approximations to the
ground--state energy in strongly correlated situations. Allowing the
functional to depend on the projection of the Kohn--Sham energies onto a
given orbital, allows the possibility of orbitally ordered states. \ This is
a major advance over LDA in situations where this orbital order is present.
As recognized many years ago, this is a very efficient way of gaining energy
in correlated situations, and is realized in a wide variety of systems.

Those are the formal difficulties of the LDA+U method. From a \ practical
point of view, despite the great successes of the LDA+U theory in predicting
materials properties of correlated solids (for a review, see Ref. \onlinecite
{Anisimov}) there are obvious problems of this approach when applied to
metals or to systems where the orbital symmetries are not broken. The most
noticeable is that it only describes spectra composed of Hubbard bands. We
have argued in the previous sections that a correct treatment of the
electronic structure of strongly correlated electron systems has to treat
both Hubbard bands and quasiparticle bands on the same footing. Another
problem occurs in the paramagnetic phase of Mott insulators, in the absence
of any broken symmetry the LDA + U method reduces to the LDA, and the gap
collapses. In systems like NiO where the gap is of the order of eV, but the
Neel temperature is a few hundred Kelvin, it is unphysical to assume that
the gap and the magnetic ordering are related. For this reason the LDA+U
predicts magnetic order in cases that it is not observed, as, e.g., in the
case of Pu \cite{Pu-LDA+U}.

\section{Model Hamiltonians and First Principles Calculations}

It is useful to think of a program for performing realistic electronic
structure calculations for correlated materials in the light of the
qualitative discussion of the earlier section. The hamiltonian describing
electrons at short distances is known and easily written down. This is the
formal starting point of all--electron first--principles calculations. So,
the theory of everything is summarized in the action $S$ of a system of
electrons with Coulomb interactions between them, described in
by the action of eq. (\ref{iham}).

This action ignores relativistic effects, which are important for the
structural properties of heavy elements. Relativistic corrections which
introduce spin--orbit interaction have the form $%
H_{FS}=H_{SO}+H_{SOO}+H_{SS} $

\[
H_{SO}=\alpha ^{2}\frac{Z}{2}\sum_{i,j}\frac{1}{r_{ij}^{3}}s_{i\cdot }l_{j} 
\]
\[
H_{SOO}=-\frac{\alpha ^{2}}{2}\sum_{i<j}\frac{1}{r_{ij}^{3}}r_{ij}\times
p_{i}(s_{i}+2s_{j}) 
\]
\[
H_{SS}=\frac{\alpha ^{2}}{2}\sum_{i<j}\frac{1}{r_{ij}^{3}}%
[s_{i}.s_{j}-(r_{ij}.s_{i})\frac{(r_{ij}.s_{j})}{r_{ij}^{2}}] 
\]
In addition this action should be supplemented by the electron--phonon
interaction which can be not negligible in correlated materials. They have
been investigated using DMFT within model calculations\cite{freericks}, but
realistic studies are in its infancy and will not be considered here.

Introduction of a complete tight binding basis $\phi _{l}({\bf x})$ allows
us to rewrite the action in the form 
\begin{equation}
S=\int d\tau [ {c_{\alpha }}^{\dagger }O_{\alpha ,\beta }\partial _{\tau
}c_{\beta }+H[{c_{\alpha }}^{\dagger }c_{\beta }]]
\end{equation}
with the Hamiltonian containing an infinite number of bands, 
\begin{equation}
H=-\sum_{ij}\sum_{lm}\sum_{\sigma }t_{ij}^{lm}[c_{i\sigma }^{+l}c_{j\sigma
}^{m}+h.c.]+\sum_{iji^{\prime }j^{\prime }}\sum_{klmn}\sum_{\sigma \sigma
^{\prime }}V_{iji^{\prime }j^{\prime },\sigma \sigma ^{\prime
}}^{klmn}c_{i\sigma }^{+k}c_{j\sigma ^{\prime }}^{+l}c_{i^{\prime }\sigma
}^{m}c_{j^{\prime }\sigma ^{\prime }}^{n}  \label{tb}
\end{equation}
the new operators $c$ are related to the continuum operators $\psi $ by $%
c_{i\sigma }^{l}=\int_{{\bf x}}\psi _{\sigma }({\bf x})\phi _{l}({\bf x}-%
{\bf R}_{i})$. $k,l,m,n$ denote band indices, $\sigma \sigma ^{\prime }$ the
spin indices and $ij$ denote the lattice sites. O describes the overlap
matrix if this basis is not orthogonal.

However, to describe the physics at a lower energy scale one would like to
eliminate the degrees of freedom which have energies much larger than that
scale, and derive an effective Hamiltonian which is more transparent and
contains only the relevant or active degrees of freedom. The effective
Hamiltonian at that scale, is the model Hamiltonian which is usually written
down by the solid--state physicist on physical grounds.

The explicit construction can be written down as a Wilsonian elimination of
the irrelevant high energy degrees of freedom. Formally, one divides the set
of operators in the path integral in $c_{H}$ describing the ''high energy ''
orbitals, and $c_{L}$ the low--energy orbitals that one would like to
consider explicitly. 
\begin{equation}
exp(-{S_{eff}[c_{L}^{^{\dagger }}c_{L}])}=\int Dc_{H}^{^{\dagger
}}Dc_{H}exp-S[c_{H}^{^{\dagger }},c_{L}^{^{\dagger }},c_{L},c_{H}]
\label{rg}
\end{equation}
In the electronic--structure program this procedure is called downfolding
and is carried out at the level of Kohn--Sham orbitals. The transformation (%
\ref{rg}) generates retarded interactions of arbitrarily high order. However
there are good reasons, why an approximation in which only quartic terms are
kept, and its frequency dependence ignored is an excellent approximation.
This results in models of the form (\ref{tb}) except that the interactions
and the hoppings are screened relative to their bare values.

If one keeps the constant part of the effective action, which contains the
free energy of the high--energy orbitals which have been eliminated this
procedure in principle contains all the information which is needed to carry
out total energy calculations. Nothing is lost except for the higher--order
interactions and the retardation which is unlikely to be very important. The
hamiltonians obtained in this way are the model hamiltonians used in
many--body theory.

Finally, the elimination in Eq (\ref{rg}) cannot be performed even within
the approximations cited above. However any technique which can be used to
treat Hamiltonians approximately, can also be used to performed the
elimination (\ref{rg}). In particular carrying the local density
approximation has been applied to obtain the $U$'s. It would be interesting
to reconsider the validity of these approximations from the point of view of
an effective action.

As the number of orbitals and the relevant energy scale is reduced a
renormalization group (RG) flow in the space of all hamiltonians is defined.
Different initial conditions at short distances describe different
substance, or materials at different pressures, lattice spacings, dopant
concentration etc.

If one starts with conditions that correspond to weakly correlated systems (
e.g. atomic numbers involving s or p electrons, high densities etc. ) the RG
flows are relatively simple and converge at low energies to reach simple
fixed points describing band metals or insulators.

On the other hand when we start from more correlated situations (e.g. open
shells, containing relatively localized $d$ or $f$ electrons, lower
densities), the RG trajectories are diverging from one another, reflecting
the diversity of phases nearby. This situation calls for quantitative
methods for realistic modelling of the material in question. One of the most
serious difficulties in carrying out the Wilson RG program described above,
is the continuous change in the {\it form} of the effective Hamiltonian from
scale to scale. A typical example is the formation of a heavy fermion liquid
state at a coherence energy scale. At high energies the effective
Hamiltonian contains atomic configurations and conduction electrons, at low
frequencies only heavy quasiparticles are the relevant degrees of freedom.
In spite of these difficulties, an R.G analysis taking account some quantum
chemistry in the initial conditions has been carried out {\it in the local
approximation} aided by developments in DMFT\cite{sikotliar}

While following the R.G. flows down to very low temperatures and predicting
physical properties in the most strongly correlated situations may prove to
be very difficult, there are many reasons to believe that C--DMFT with small
sizes will be accurate in a wide range of interesting situations (not too
close to phase transitions, not too low temperatures).

The process of eliminating degrees of freedom with the approximations
described above gives us a physically rigorous way of thinking about
effective Hamiltonians with effective parameters which are screened by the
degrees of freedom to be eliminated. Since we neglect retardation, and terms
of higher order than six the effective hamiltonian would have the form

\begin{equation}  
H_{eff}=-\sum_{ij}\sum_{lm}\sum_{\sigma }t_{ij}^{lm}[c_{i\sigma
}^{+l}c_{j\sigma }^{m}+h.c.]+\sum_{iji^{\prime }j^{\prime
}}\sum_{klmn}\sum_{\sigma \sigma ^{\prime }}V_{iji^{\prime }j^{\prime
},\sigma \sigma ^{\prime }}^{klmn}c_{i\sigma ^{\prime }}^{+k}c_{j\sigma
}^{+l}c_{i^{\prime }\sigma }^{m}c_{j^{\prime }\sigma ^{\prime }}^{n}+E_{0}
\label{tb1}
\end{equation}
It should be regarded as the effective hamiltonian \ that one can use to
treat the relevant degrees of freedom. If the dependence of $E_{0}$ on the
nuclear coordinates are kept, it can be used to obtained the total energy.
If the interaction matrix turns out to be short ranged or has a simple form,
this effective Hamiltonian could be identified with the Hubbard or the
Anderson hamiltonians which have been treated in the literature.

To conclude this section it is worth clarifying several terms that are used
in electronic--structure literature, the light of the previous discussion,
for pedagogical reasons. The first point is the meaning of {\it ab initio}
or first--principles calculations. These imply that no empirically adjus
table parameters are needed in order to predict physical properties of
compounds, except the structure and the charges of atoms. First principles
do not mean exact or accurate or computational inexpensive. If the effective
hamiltonian is derived, i.e. if the functional integral is performed by a
set of well--defined approximations, and the consequent hamiltonian (\ref
{tb1}) is solved for its total energy keeping track of the constant $E_{0}$
we have a first--principles method. In practice, the derivation of the
effective hamiltonian or its solution may be inaccurate or impractical in
which case the {\it ab initio} method is not very useful.\ We make this
remark because $H_{eff}$ has the form of \ a "model Hamiltonian" and very
often a dichotomy between model hamiltonians and first--principles
calculations is made. What makes a model calculation semiempirical is the
lack of coherent derivation of the form of the "model hamiltonian" and the
parameters entering there. A second point has to do with the elimination of
degrees of freedom and the meaning of \ all--electron calculation,
indicating that all the electronic degrees of freedom are taken into
account. All electronic structure methods make some elimination of degrees
of freedom, it is most evident in pseudopotential methods, but it is also
true in other methods. To make a problem in the continuum computationally
feasible, a discretization is necessary. The elimination of unnecessary
degrees of freedom greatly facilitates the accuracy of the consequent
discretization. Introduction of a finite basis set in electronic--structure
calculations is a form of discretization, the important question is its
accuracy. The LMTO's have proved to be extraordinarily accurate to
discretize the Kohn--Sham hamiltonian in the relevant region of energy. In
DMFT not only the one--body term of the Hamiltonian is truncated, but also
the interaction terms are, this truncation\ is most accurate in
ultralocalized basis. The C--DMFT and the E--DMFT are many--body approaches
which lend themselves to practical truncations in non--orthogonal LMTO's.

\section{Dynamical mean--field theory}

>From a conception point of view, the construction of dynamical mean--field
functional constitutes a radical departure from Kohn--Sham based DFT. The
dynamical mean--field equations as we will see are the equations for a
continuous distribution of spectral weight. Attention has been shifted away
from the well--defined Kohn--Sham quasiparticles (poles in the Green
function) to continuous distribution of spectra (which appear as branch
cuts) and give rise for example to the Hubbard bands.

Another central difference connected to this point is the fact that the
local spectral function can now be identified with the observable
one--electron spectrum. This is very different from the Kohn--Sham
quasiparticles which cannot be identified rigorously with the one--electron
spectra. While DMFT is computationally more demanding than DFT, it is
formulated in terms of observables and gives more information than DFT. It
can be formulated in an effective action point of view in complete analogy
with DFT.

Our starting point is a multiband Hubbard Hamiltonian,$i$ $j$ denote lattice
sites while

$\alpha $ $\beta $ are spin -orbital indices:

\begin{equation}
H=\sum_{ij}\sum_{\alpha \beta }t_{ij}^{\alpha \beta }c_{i\alpha
}^{+}c_{j\beta }+\sum_{ij}\sum_{\alpha \beta \gamma \delta }V_{ij}^{\alpha
\beta \gamma \delta }c_{i\alpha }^{+}c_{i\beta }^{+}c_{j\delta }c_{j\gamma }
\label{start}
\end{equation}
which is obtained from the procedure discussed in the previous section. The
effective action construction of \ DMFT parallels that given in other
sections. A source $J_{i\alpha \beta }({\bf \tau },{\bf \tau }^{\prime })$
is introduced in the partition function

\begin{equation}
e^{-\beta W[j]}=\int dc^{+}dce^{-S-\sum_{i\alpha \beta }\int J_{i\alpha
\beta }({\bf \tau },{\bf \tau }^{\prime })c_{i\alpha }^{+}({\bf \tau }%
)c_{i\beta }({\bf \tau }^{\prime })d{\bf \tau }d{\bf \tau }^{\prime }}
\label{part1}
\end{equation}
$W$ generates the local Green function 
\begin{equation}
\frac{\delta W}{\delta j_{i\alpha \beta }({\bf \tau },{\bf \tau }^{\prime })}%
=-\left\langle c_{i\alpha }({\bf \tau })c_{i\beta }^{+}({\bf \tau }^{\prime
})\right\rangle =A_{\alpha \beta }({\bf \tau },{\bf \tau }^{\prime })
\label{green1}
\end{equation}
The effective action is $\Gamma (A)=W[j(A)]-J(A)A.$ The minimum 
\begin{equation}
\frac{\delta \Gamma }{\delta A}=0  \label{deltagamma}
\end{equation}
gives rise to correct spectra and total energy. Notice that the definition
of the local Green function depends on the basis of orbitals chosen, this is
similar to the LDA+U method. However, unlike the LDA+U\ approach, we are
dealing with the functionals of Green functions which have a meaning
independently of the Kohn--Sham representation of the density.

Notice that the definition of the local Green function requires the choice
of a basis. However, unlike the LDA+U method, the choice of basis is used to
perform a Legendre transfomation with respect to a well--defined object, the
local Green function, rather than with respect to a part of the density, as
in the LDA+U\ method which lacks a clear physical significance.

The starting point (\ref{start}) can be regarded as a model hamiltonian,
but, as argued in the previous section, if the constant parts are kept and (%
\ref{tb1}) is carefully derived, this is effectively equivalent to the full
hamiltonian in the relevant energy range.

The dynamical mean--field approximation to the functional $\Gamma $, i.e. $%
\Gamma _{DMFT},$ can be written in two alternative forms depending on
whether we stress that it is a truncation of the exact functional when
expanding $\Gamma $ in powers of the hopping (atomic expansion) or in powers
of the interaction (expansion around the band limit). To write this
functional it is useful to define the quantity $\chi (i\omega _{n},A).$The
physical meaning of this expression is parallel to the meaning of the
Kohn--Sham potential: it is the function that one needs to add to the free
hamiltonian in order to obtain a desired spectral function, as described in
table I.

\begin{center}
\large
\begin{table}[tbp]
\caption{ Parallel between the different approaches, indicating the physical
quantity which has to be extremized, and the field which is introduced to
impose a constraint (Kohn Sham field).}\centering
\par
\begin{tabular}{lll}
Method & Physical Quantity & Constraining Field \\ \hline
LDA & $\rho (r)$ & $V_{KS}(r)$ \\ 
DMFT (band limit) & $A_{\alpha \beta }(i\omega )$ & $\chi _{\alpha \beta
}(i\omega )$ \\ 
DMFT (atomic limit) & $A_{\alpha \beta }(i\omega )$ & $\Delta _{\alpha \beta
}(i\omega )$ \\ 
Baym-Kadanoff & $G_{\alpha \beta }({\bf k},i\omega )$ & $\Sigma _{\alpha
\beta }({\bf k},i\omega )$ \\ 
LDA+U & $\rho (r),n_{ab}$ & $V_{KS}(r),\lambda _{ab}$ \\ 
LDA+DMFT (band limit) & $\rho (r),A_{\alpha \beta }(i\omega )$ & $%
V_{KS}(r),\chi _{\alpha \beta }(i\omega )$ \\ 
LDA+DMFT (atomic limit) & $\rho (r),A_{\alpha \beta }(i\omega )$ & $%
V_{KS}(r),\Delta _{\alpha \beta }(i\omega )$%
\end{tabular}
\end{table}
\end{center}

\begin{equation}
A^{\alpha \beta }(i\omega _{n})=\sum_{{\bf k}}[i\omega _{n}-t({\bf k}%
)-\epsilon _{0}-\chi (i\omega _{n},A)]_{\alpha \beta }^{-1}  \label{scdmf}
\end{equation}
This quantity is important to set up an expansion of the functional in
powers of the interaction following the inversion method. This emphasizes
DMFT as a partial summation of the expansion around the band limit. Notice
that if the exact self--energy of the problem is ${\bf k}$ independent, then 
$\chi $ coincides with the self--energy, this statement is parallel to the
observation within DFT: if the self--energy of a model is ${\bf k}$ and
frequency independent then the self--energy coincides with the Kohn--Sham
potential. The DMFT functional is given by

\begin{equation}
\Gamma _{DMFT}(A,\chi )=-\sum_{{\bf k}}\sum_{i\omega _{n}}tr\log (i\omega
_{n}-t({\bf k})-\epsilon _{0}-\chi )-\sum_{i\omega _{n}}tr\chi A(i\omega
_{n})+\Phi \lbrack A]  \label{gammaachi}
\end{equation}
where $\Phi \lbrack A]$ is a sum of all local graphs (on a single site i)
constructed with $V_{iiii}^{\alpha \beta \gamma \delta }$ as a vertex and $A$
as a propagator which are two-particle irreducible. The diagrammatic rules
for the exact functional are more complicated and were discussed in ref. [ 
\onlinecite
{chitrafunc1}].

Eq. (\ref{scdmf})\ \ determining $\chi =\chi (A)$ appears as a saddle point
of the functional (\ref{gammaachi}) and should be \ backsubstituted to
obtain $\Gamma _{DMFT}(A)$, the DMFT approximation to $\Gamma (A).$

The second derivation emphasizes the expansion around the atomic limit,
where the starting point is a dressed atom. It starts by\ writing the
hamiltonian (\ref{tb1}) into two parts: $H=H_{0}+H_{1}$ with $H_{0}=\sum
h_{0}[i]$, where 
\begin{equation}
h_{0}[i]=-\sum_{lm}t_{ii}^{lm}[c_{i\sigma }^{+l}c_{i\sigma
}^{m}+h.c.]+\sum_{klmn}V_{iiii\sigma \sigma ^{\prime }}^{klmn}c_{i\sigma
^{\prime }}^{+k}c_{i\sigma }^{+l}c_{i\sigma }^{m}c_{i\sigma ^{\prime }}^{n}
\label{h0sc}
\end{equation}
and $H_{1}=H-H_{0}$ and carries out the inversion method in powers of $%
\lambda H_{1}$ ($\lambda $ is a coupling constant to be set to unity at the
end). Here the indices k, l, m,n are orbital indices.

The zeroth--order term requires the introduction of the Kohn--Sham field, $%
J_{0}\equiv J_{0{lm}}^{\alpha \beta },$ which couples to the local lattice
Green's function $A_{lm}^{{}}$ The effective action involving $H_{0}$ and $%
J_{0}$ has the form of an impurity model where the source lays the role of
the hybridization \ therefore we will switch notation and sometimes use $%
\Delta (\tau ,\tau ^{\prime })$ for this source. This action is highly
non--trivial due to the presence of the impurity--interaction term in (\ref
{h0sc}). The lowest--order term in the inversion method then requires us to
solve an impurity model which expresses the source $\Delta (\tau ,\tau
^{\prime })$ in terms of the impurity model Green functions of a generalized
Anderson impurity model defined by 
\begin{equation}
S_{at}[\Delta ]=\int d\tau d\tau ^{\prime }\sum_{\sigma lm}c_{\sigma
}^{+l}(\tau )\left[ \delta (\tau -\ \tau ^{\prime }){\frac{{\partial }}{{%
\partial \tau ^{\prime }}}}+\Delta ^{lm}(\tau -\tau ^{\prime })\right]
c_{\sigma }^{m}(\tau ^{\prime })+\int d\tau H_{0}  \label{scact}
\end{equation}
The impurity Green function is given by $A_{lm}^{\alpha \beta }(\tau ,\tau
^{\prime })$

\begin{equation}
\frac{\delta W_{at}}{\delta J_{lm}({\bf \tau },{\bf \tau }^{\prime })}%
=-\left\langle c_{l\sigma }({\bf \tau })c_{m\sigma }^{+}({\bf \tau }^{\prime
})\right\rangle _{j_{0}}=A_{lm}^{{}}(\tau ,\tau ^{\prime })  \label{shamatom}
\end{equation}
where the expectation value of any operator $O$ is given by 
\begin{equation}
\langle O\rangle _{J_{0}}={\frac{{\int Dc_{\sigma }^{+l}Dc_{\sigma
}^{m}O\exp -S[\Delta ]]}}{{\int Dc_{\sigma }^{+l}Dc_{\sigma }^{m}\exp
-S[\Delta ]}}}  \label{41}
\end{equation}
and 
\[
W_{at}[\Delta ]=-\log \int dc^{+}dce^{-S_{at}[c^{+}c]} 
\]
describes an atom or a set of atoms in the unit cell embedded into the
medium., $\alpha $ labels spin, and orbital including position in the unit
cell. $\Delta $ is the Weiss field of the mean--field theory. It is simply
the Kohn--Sham field (with respect to the expansion around atomic limit)
which is defined by the equation expressing that $\Delta $ gives rise to the
local Green function $A$ (see Eq. (\ref{shamatom}) where the general
Kohn--Sham source was identified with the hybridization of the Anderson
impurity model).

The functional corresponding to (\ref{scact}) is now given by 
\begin{eqnarray}
\Gamma _{0}[A] &=&W_{at}[\Delta {\lbrack A]}]-\Delta {\lbrack A]}A
\label{func} \\
&=&Tr\log A-A_{at}^{-1}A+\Phi \lbrack A]  \nonumber
\end{eqnarray}
with the $A_{at}^{-1}=i\omega _{n}-{\bar{\epsilon}}$ where ${\bar{\epsilon}}%
=\sum_{{\bf k}}\epsilon ({\bf k})$ ($\epsilon ({\bf k})$ is the Fourier
transform of the hoppings $t_{ij}$, $W_{at}$ the free energy $\lambda =0$
and $\Phi $ is the sum of all one particle irreducible diagrams constructed
with the local vertex $V_{iiii}$ and $A$. 

Now the functional $\Gamma \lbrack A]$ can in principle be constructed if
all terms in the expansion in $\lambda $ could be summed up. The DMFT
approximation is obtained by summing all diagrams with a topology of a
cactus, describing \ multiple excursion away from a given site, leading to
the expression 
\begin{eqnarray}
&&\Gamma _{DMFT}(A,\chi ,\Delta )=W_{at}[\Delta ]-\sum_{i\omega
_{n}}tr(\Delta A)-\sum_{i\omega _{n}}tr\log A-  \label{functional} \\
&&\frac{1}{N_{s}}\sum_{{\bf k}}\sum_{i\omega _{n}}tr\log (i\omega _{n}-t(%
{\bf k})-\epsilon _{0}-\chi )-\sum_{i\omega _{n}}tr(\chi -i\omega
_{n}+\epsilon _{0})A  \nonumber
\end{eqnarray}
It is useful to check the equation obtained by differentiating with respect
to $A$ which results in the CPA\ condition: 
\[
i\omega _{n}-\epsilon _{0}-\Delta =A^{-1}+\chi (i\omega _{n}) 
\]
The $tr$ runs over the spin and orbital indices, $N_{s}$ is the number of
unit cells

One can rewrite this functional \ by expressing both $A$ and $\chi $ in
terms of the Weiss field $J_{0}$ (which in this formalism plays the role of
the Kohn Sham field of the effective action formalism when an expansion
around the atomic limit is carried out, as opposed to the expansion around
the band limit considered by Aliev and Fernando).

\begin{equation}
\Gamma \lbrack A[\Delta ],\chi \lbrack \Delta ]]=W_{0}[\Delta ]-Tr\log
A-\sum_{{\bf k}}\sum_{i\omega _{n}}Tr\log [i\omega _{n}-\epsilon ({\bf k}%
)-\chi ]-\sum_{i\omega _{n}}[\chi -i\omega _{n}+{\bar{\epsilon}}+\Delta ]A
\label{gammadmft}
\end{equation}
A second important point is that both the truncation of the expansion around
the atomic limit or the truncation of the graphs in powers of $V$, indicate
that $\Gamma _{DMFT}$ is going to be a poor approximation to $\Gamma (A)$
when the interactions $V$ are highly non--local. This lead to the
introduction of a hybrid method where the light electrons are treated by LDA
which contain screening effects while the heavy electrons are treated by the
dynamical mean--field methods. We will review this approach in the next
section.

Notice however that an extension of the DMFT formalism allow us to bypass
the introduction of LDA altogether as we will show in section \ref{gw}.

\section{LDA+DMFT}

We now turn to the LDA+DMFT method \cite{poteryaev,ldaplus} which has
recently been implemented in a self consistent way \cite{pluprl}. On one
hand, this approach can be viewed as a natural evolution of the LDA+U method
to eliminate some difficulties discussed at the end of Section\ V. On the
other hand, it can be viewed as a way to improve the DMFT\ approach, \ so as
to bring in more microscopic details to an approach that had been extremely
successful at the level of model Hamiltonians as described in section III.

We derive the equations following the effective action point of view \cite
{chitrafunc1}.To facilitate the comparaison between the approaches discussed
in the earlier sections we have tabulated the central quantities which have
to be minimized, and the fields which are introduced to impose a constraint
in the effective action method \cite{fukuda}. As in the LDA+U method one
introduces a set of correlated orbitals $\phi _{a}({\bf r}-{\bf R}).$ One
defines an exact functional of the total density $\rho ({\bf r})$ and of the
local spectral function of the correlated orbitals discussed before: 
\begin{equation}
A_{ab}(i\omega ,R)=-\int \phi _{a}^{\ast }({\bf r-}R)\left\langle \psi ({\bf %
r},i\omega )\psi ^{+}({\bf r}^{\prime },i\omega )\right\rangle \phi _{b}(%
{\bf r}^{\prime }-R)d{\bf r}d{\bf r}^{\prime }=-\left\langle c_{a}(i\omega
,R)c_{b}^{+}(i\omega ,R)\right\rangle  \label{Aab}
\end{equation}
where indexes $a,b$ refer exclusively to the heavy orbitals.

We now introduce sources for the density and the local spectral function of
the heavy orbitals, $l({\bf r})$, and $j_{ab}({\bf R,\tau },{\bf \tau }%
^{\prime })$\ 

\begin{equation}
e^{-\beta W[l,j]}=\int dc^{+}dce^{-S-\int l({\bf r})\rho ({\bf r})d{\bf r}%
-\sum_{abR}\int j_{ab}(R,{\bf \tau ,\tau }^{\prime })c_{aR}^{+}({\bf \tau }%
)c_{bR}({\bf \tau }^{\prime })d{\bf \tau }d{\bf \tau }^{\prime }}
\label{embw}
\end{equation}
\begin{equation}
\frac{\delta W}{\delta l({\bf r})}=\rho ({\bf r})  \label{dwdl}
\end{equation}
\begin{equation}
\frac{\delta W}{\delta j_{ab}(R,{\bf \tau },{\bf \tau }^{\prime })\ }%
=-\left\langle c_{Ra}({\bf \tau )}c_{Rb}^{+}({\bf \tau }^{\prime
})\right\rangle =A_{ab}({\bf \tau },{\bf \tau }^{\prime })\   \label{dwdj}
\end{equation}
Then, the functional of both density and the spectral function is
constructed by Legendre transform. This is an exact functional of the
density and the local\ Green function, $\Gamma (\rho ,A)$, which gives the
energy at the stationary point, and for which in principle, \ a perturbative
construction can be carried either around the atomic limit or around the
band limit following the inversion method.

However, based on the evidence that even at the level of simple model
hamiltonians DMFT describes accurately the properties of many systems and
contains physics which is not captured by any other approach, and based on
the remarkable success of LDA in treating weakly correlated electron
systems, a useful approximation suggests itself, \ in a form of a LDA+DMFT\
approximation.

The functional implementation corresponding to this approximation is given
by $\Gamma _{LDA+DMFT\;}(\rho ,V_{KS,}\Sigma ,A)$ which has the form 
\begin{eqnarray}
\Gamma _{LDA+DMFT}(\rho ,V_{KS,}\Sigma ,A) &=&-T\sum_{i\omega _{n}}tr\log
[i\omega _{n}+\nabla ^{2}/2-V_{KS}-\Sigma _{ab}(i\omega _{n})\phi _{a}({\bf r%
})\phi _{b}^{\ast }({\bf r}^{\prime })]-  \label{gammalda+dmft} \\
&&\int V_{KS}({\bf r})\rho ({\bf r})d{\bf r}-\sum_{i\omega
_{n}}\sum_{ab}\Sigma _{ab}(i\omega _{n})A_{ba}(i\omega _{n})+  \nonumber \\
&&\int V_{ext}({\bf r})\rho ({\bf r})d{\bf r}+\frac{1}{2}\int \frac{\rho (%
{\bf r})\rho ({\bf r}^{\prime })}{|{\bf r}-{\bf r}^{\prime }|}d{\bf r}d{\bf r%
}^{\prime }+E_{xc}^{LDA}[\rho ]+  \nonumber \\
&&\sum_{j}\Phi \lbrack A_{ab}(j,i\omega )]+\Phi _{DC}  \nonumber
\end{eqnarray}

The field $\chi$ of the previous section is now denoted by $\Sigma$.
Together with $V_{KS}$ they are the fields needed to constrain the density
and the local spectral function of their correlated orbitals to their given
values $\rho$ and $A$. $\Phi $ is the sum of all two--particle irreducible
graphs constructed with the local part of the interaction \ \ and $\Phi
_{DC} $ is taken to have the same form as in LDA+U method, \ i.e. a simple
local Hartree--Fock form with $n_{ab}=T\sum A_{ab}(i\omega )e^{i\omega
_{n}0^{+}}$As we argued before in a fixed tight--binding basis, $-\nabla
^{2}+V_{KS}$ reduces to $H^{TB}({\bf k}) $ and the functional $\Gamma
_{LDA+DMFT}$ \ for a fixed density and truncated to a finite basis set takes
a form identical to the DMFT\ functional Eq. (\ref{gammadmft}) in the
section VII. Its minimization leads to the set of equations with Kohn--Sham
potential defined by Eq. (\ref{vks}) and 
\begin{equation}
\Sigma _{ab}(i\omega _{n})=\frac{\delta \Phi }{\delta A_{ab}(i\omega _{n})}%
+\epsilon _{ab}^{DC}  \label{sigmaabw}
\end{equation}
which identifies $\Sigma $ as a self--energy of a generalized Anderson
impurity model in a bath characterized by a matrix of levels 
\begin{equation}
\epsilon _{ab}^{0}=\epsilon _{ab}^{DC}+\sum_{{\bf k}}H_{ab}^{TB}({\bf k})
\label{eab0}
\end{equation}
and a hybridization function $\Delta _{ab}(i\omega _{n})$ obeying a
self--consistency condition

\begin{equation}
i\omega O_{ab}-\epsilon _{ab}^{0}-\Delta _{ab}(i\omega _{n})=\Sigma
_{ab}(i\omega _{n})+\left[ \sum_{k}(i\omega _{n}O-\epsilon ^{0}-t({\bf k}%
)-\Sigma (i\omega _{n}))^{-1}\right] _{ab}{}^{-1}  \label{deltaabw}
\end{equation}
Finally, minimizing Eq. (\ref{gammalda+dmft}) with respect to $V_{KS}$
indicates that $\rho ({\bf r})$ should be computed as

\begin{equation}
\rho ({\bf r})=T\sum_{i\omega _{n}}\left\langle {\bf r}\left| [i\omega
_{n}+\nabla ^{2}/2-V_{KS}-\sum_{ab}\Sigma _{ab}(i\omega _{n})\phi _{a}({\bf r%
})\phi _{b}^{\ast }({\bf r}^{\prime })]^{-1}\right| {\bf r}\right\rangle
e^{i\omega _{n}0^{+}}  \label{rholda+dmft}
\end{equation}
which as indicated before, when truncating in a fixed set of orbitals
becomes 
\begin{equation}
\rho ({\bf r})=T\sum_{i\omega _{n}}\sum_{\alpha \beta }\chi _{\alpha }^{\ast
}({\bf r})\left[ (i\omega _{n}-H^{TB}({\bf k})-\Sigma (i\omega _{n})\right]
^{-1}\chi _{\beta }^{\ast }({\bf r})  \label{rhoagain}
\end{equation}

The solution of Eqs.(\ref{sigmaabw})--(\ref{rhoagain}) is carried out in
double iterational loop described in Fig.~7. It is important to
note that the choice of basis $\chi _{\alpha }^{{\bf k}}$ determines the
quality of the truncation \ in a given energy range.

\begin{figure}[htb]
\centerline{\epsfig{file=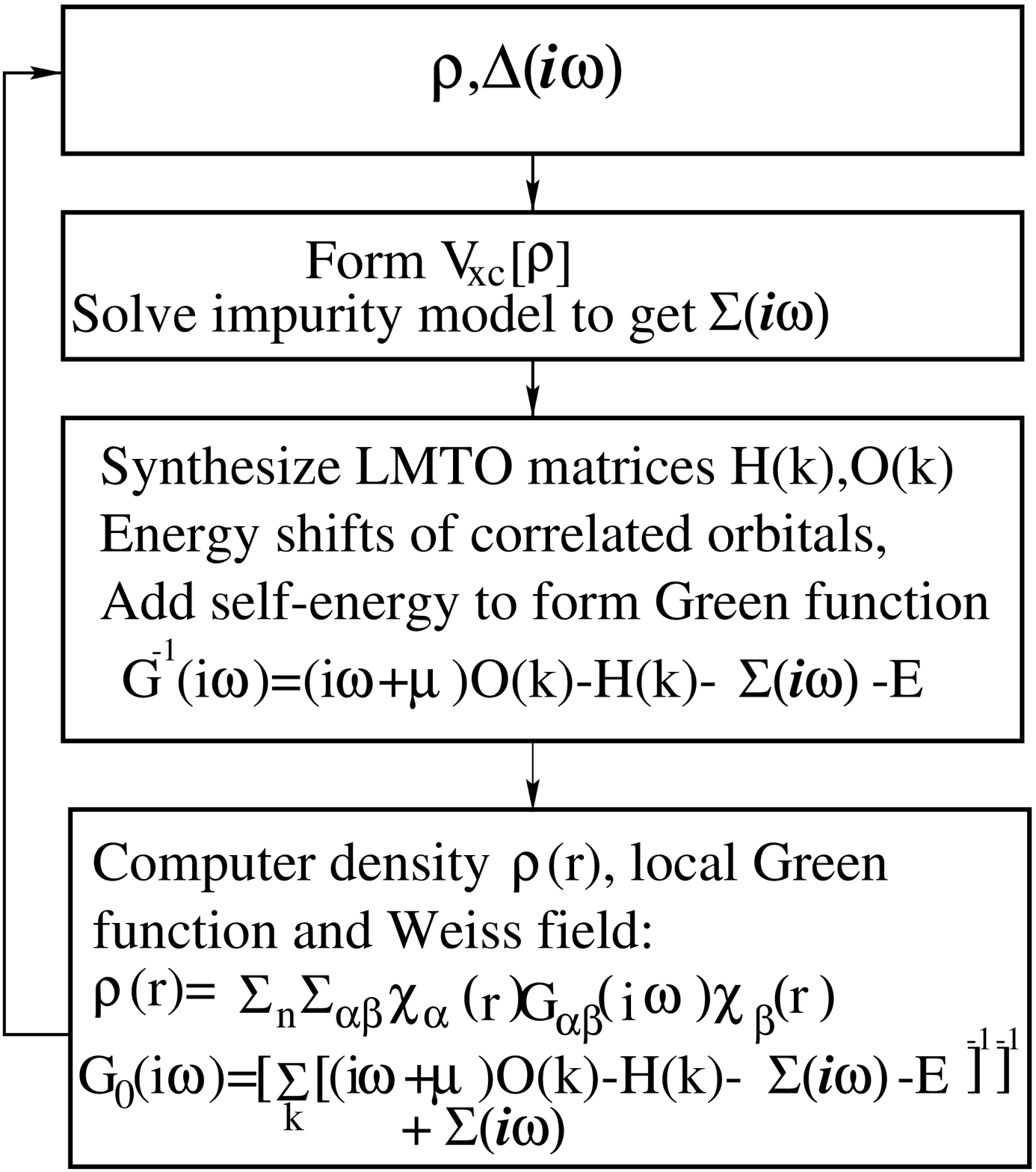, height=16.0cm}}
{FIG~7. Schematic description of the implementation of the LDA+DMFT loop
carried out in ref \protect\onlinecite{savrasov}}
\end{figure}


The basis can be gradually refined so as to obtain more accurate solutions
in certain energy range. In principle this improvement is done by changing
the linearization energies, and the experience from density functional
implementations could be carried over to the DMFT case. Finally notice that
the\ rational for the counter term described in Eq. (\ref{eab0}) originates
in comparison of spectra with the results of LDA+U calculations and deserves
further investigations.

A different possibility which is computationally very practical, is to adopt
the\ philosophy of the generalized IPT, introduced earlier\cite
{kajueterkotliar} and to fix the double counting term by requesting the
fulfillment of the Luttinger theorem. i.e. to chose

\begin{equation}
\epsilon _{ab}^{DC}=-\Sigma _{ab}(0)  \label{eabdc}
\end{equation}

A version of \ this counter term, has been recently applied to iron with
encouraging results \cite{sashafe} The ambiguity in the choice of double
counting correction may turn out to be hard to resolve since one is
substracting terms from a non diagrammatically controlled approximation. The
\ ideas described in section XII, which \ results in a full Green function
treatment of the many-- body problem is perhaps the most promising route
towards a non empirical resolution of this ambiguity.

\section{Techniques for solving realistic DMFT equations}

In practice the solution of the DMFT equations is more involved than the
solution of the KS equations, which now appear as static analogs. Even for
the simplest model hamiltonians, some care and judicious choice is required
to obtain correct results. There are two central aspects of DMFT. One is the
self--consistency condition, Eq. (\ref{deltaabw}). This step is trivial for
model calculations but becomes time consuming when realistic bands are
introduced. It is done using tetrahedron methods. Second, one has to solve
the impurity model. Fortunately, we can now rely on several years of
experience to device reasonable approximations. First, at sufficiently high
temperature QMC calculations are possible. Notice however the poor scaling
with orbital degeneracy. The approach that we have advocated to carry out
this step is very similar in spirit to the analytic parametrizations of $%
V_{xc}$ in LDA. One uses different approximations to the self--energy of the
impurity model viewed as a functional of $\Delta (i\omega )$ in different
regions of frequency. 
\cite{GeneralizedIPT}.

\section{Successes}

The early studies on V$_{2}$O$_{3}$ and NiSeS which were qualitative in
nature, indicated the great potential of the DMFT\ approach. Another early
contribution was a series of papers which display semiquantitative agreement
on the 30\% level \ on a series of bulk physical properties for the LaSrTiO
system. The specific heat was considered\cite{marceloprb}, the magnetic
susceptibility was computed, the optical conductivity\ was calculated\cite
{marcelo1}, the Hall effect was obtained\cite{henrik0}. Finally the
thermoelectric power was focused\cite{henrik1}. On those fillings, the
difference between the one band and the multiband situation for low--energy
properties was estimated not to be very large for low energy properties
given the uncertainties in the determination of the interaction parameters.
Given the fact that only very simple tight--binding parametrizations were
used in those works, and the fact that a large number of experiments were
fit with the same value of parameters\ one should regard the agreement as
satisfactory. The photoemission spectroscopy of this compound, as well as
others, are not completely consistent with the bulk data, and several
authors have argued that disorder and modelling of the specific surface
environment is required to improve the agreement with experiment\cite{sarma}.

Given excellent agreement at the level of model hamiltonian calculations it
was natural to incorporate the full band structure into DMFT to perform
calculations of spectra which would incorporate more high energy physics,
such as additional bands. This was first implemented\cite{poteryaev,ldaplus}

In all these systems the LDA+DMFT\ approach gives results which are in
better agreement with experiments than simple LDA calculations. We have
recently argued argued that correlation effects are important for subtle
effects such as the magnetic anisotropy. The easy axis of iron and Ni were
obtained correctly\cite{imseok}. Lichtenstein and collaborators have
obtained an excellent theoretical description of the spectra of iron and
nickel\cite{lichtenstein}.

A major recent development is the\ full self--consistent implementation of
the DMFT+LDA approach. This development sheds a light on the mysterious
properties of $\delta $ Plutonium, and has resulted in new physical picture
of the $\alpha $ phase of this material\cite{savrasov2}. A second recent
development is the development of the algoritms to perform transport
coefficients which allowed the evaluation of thermoelectrical properties of
correlated materials\cite{palssonphd}.

\section{ Connection with Landau functional}

It is useful to consider explicitly the limiting case, of a single band
model on a lattice, and to evaluate the functional (\ref{functional})\ in
the simple case of the semicircular density of states with half bandwidth D,
a toy model which \ has been studied intensively over the past few years. In
this case Eq. (\ref{scdmf}) which expresses $\chi ,$or $\ z\equiv i\omega
-\varepsilon -\chi $in terms of A, becomes $A=\frac{2}{D^{2}}[z-\sqrt{%
z^{2}-D^{2}}]$which is easily inverted \ to give $z=t^{2}A+A^{-1}$with $2t=D$%
.

The energy functional can be evaluated explicitly since $\sum \log [z-t_{k}]=%
\frac{z}{D^{2}}[z-\sqrt{z^{2}-D^{2}}]+Log[\frac{[z+\sqrt{z^{2}-D^{2}}]}{2}]+%
\frac{1}{2}$and reduces to

\[
{W_{imp}[\Delta ]+}T\sum_{\omega }(zA-2\Delta (i\omega )A-1)={W_{imp}[\Delta
]+}T\sum_{\omega }(t^{{}}A-\frac{1}{t}\Delta (i\omega ))^{2}-\frac{\Delta
(i\omega )^{2}}{t^{2}} 
\]
where 
\begin{equation}
{-\beta W_{imp}}=Log\int df^{+}dfe^{-L_{loc}[f^{+},f]-{\sum_{\omega ,\sigma }%
}{f^{+}}_{\sigma }(i\omega )\Delta (i\omega ){f}_{\sigma }(i\omega )}
\end{equation}
and $L_{loc}[f^{+},f]={{\int_{0}}^{\beta }}f_{\sigma }^{+}[{{\frac{d}{{d\tau 
}}}+\varepsilon _{f}}]f_{\sigma }+U{f^{+}}_{\uparrow }f_{\uparrow }{f^{+}}%
_{\downarrow }f_{\downarrow }$.

The stationary condition which relates ${\Delta }$ to A \ is such that the
second term vanishes at stationarity, so as far as the stationary \ points
are concerned this is equivalent to the Landau functional \cite{landau}
which was considered in the context of the theory of the Mott transition.

\begin{equation}
F_{LG}{[\Delta ]}=-T\sum_{\omega }\frac{\Delta (i\omega )^{2}}{t^{2}}+F_{imp}%
{[\Delta ]}  \label{landau}
\end{equation}
This functional can be understood by analogy to the Hubbard--Stratonovich
construction of the Ising model free energy. 
\begin{equation}
\beta {F_{LG}}[h]={\beta {\frac{{h}^{2}}{{2J}}}-log[ch[2\beta h]]}
\label{strat}
\end{equation}
The first term in Eqs. (\ref{landau}) and (\ref{strat}) represents the
energetic cost of forming the Weiss field, while the second terms are the
energy gain of the local entity (spin or electron in the classical quantum
case respectively)due to the presence of the Weiss field.

It is important that the Landau functional is not just \ constructed by
Legendre transformations, but it contains additional source terms which
vanish at the saddle point. This flexibility in extending the Baym--Kadanoff
construction to obtain functionals with more desirable properties such as (%
\ref{landau}) by adding sources were discussed in Ref.[\onlinecite{chitrafunc2}].

\section{Extended DMFT, GW and GWU}

\label{gw}In this final section, we discuss how one can avoid the
introduction of LDA in our realistic DMFT framework, while still retaining
an accurate treatment of the Coulomb interactions which should allow an
accurate evaluation of the total energy. This is based on extended
dynamical--mean--field ideas, which have been developed \ in a series of
works \cite{edmft} \cite{si}. This development is particularly important to
avoid the ambiguities connected with the substraction of the
double--counting correction, discussed in section VIII.

The starting point is the Hamiltonian, 
\begin{equation}
H=-\sum_{ij}\sum_{\alpha \beta }t_{ij}^{\alpha \beta }c{^{+}}_{i\alpha
}c_{j,\beta }+\sum_{i}\sum_{\alpha \beta \gamma \delta }\Gamma ^{\alpha
\beta \delta \gamma }c_{i\alpha }^{+}c_{i\beta }^{+}c_{i\gamma }c_{i\delta
}+\sum_{ij}\sum_{ab}\sum_{\alpha \beta \gamma \delta }V_{ij}^{ab}\Lambda
_{\alpha \beta }^{a}c_{i\alpha }^{+}c_{i\beta }c_{j\gamma }^{+}c_{j\delta
}\Lambda _{\gamma \delta }^{b}
\end{equation}
The form of the Coulomb interaction is not completely general, and in
particular it omits pair hopping and correlated hopping terms of the kind
discussed by Hirsch. The corresponding action is given by 
\begin{equation}
S=\sum_{i,j}\int_{0}^{1/T}c_{i\alpha }^{+}\frac{\partial }{\partial \tau }%
O_{\alpha \beta }^{(i,j)}c_{j\beta }+\int_{0}^{1/T}Hd\tau
\end{equation}
which can be immediately transformed into an action which resembles that of
an electron--phonon problem by introducing Hubbard--Stratonovic fields for
decoupling the {\it non--local} part of the Coulomb interactions. The
physical meaning of the fields $i{\phi ^{a}}(j)$ are the electric potential
created by the multipole of the charges in all the cells different from j.
Notice that $B_{a,b}(i,i)=0$ by construction since $V_{i,i}=0$. The Bose
propagator in equation \ref{bose} simply reproduces the effects of the
Coulomb interaction. After these transformations we can apply the extended
dynamical approach to the action $S=S_{e}+S_{p}+S_{ep}$. 
\begin{equation}
{S_{p}}=\frac{1}{2}\sum_{ij}\int \int B_{ab}^{-1}(\tau i,\tau ^{\prime
}j)\phi ^{a}(\tau i)\phi ^{b}(\tau ^{\prime }j)d\tau d\tau ^{\prime }
\label{bose}
\end{equation}
\begin{equation}
{S_{e}}=\sum_{i}\sum_{\alpha \beta \gamma \delta }\Gamma ^{\alpha \beta
\delta \gamma }c_{i\alpha }^{+}c_{i\beta }^{+}c_{i\gamma }c_{i\delta
}+c_{i\alpha }^{+}\frac{\partial }{\partial \tau }O_{\alpha \beta
}^{(i,j)}c_{i\beta }-\sum_{ij}c_{i\alpha }^{+}t_{ij}^{\alpha \beta
}c_{j\beta }
\end{equation}
\begin{equation}
S_{ep}=i\sum_{i}\sum_{a\alpha \beta }\int_{0}^{1/T}\phi ^{a}(i){\Lambda }%
_{\alpha \beta }^{a}c_{i\alpha }^{+}c_{i\beta }d\tau  \label{vertex}
\end{equation}

The Baym--Kadanoff functional for this problem was derived in ref [\onlinecite
{chitrafunc2}]:

\begin{eqnarray}
\Gamma \lbrack {\bar{\phi}},G,\Pi ] &=&S[{\bar{\phi}}]+ig\sum_{i}{\Lambda }%
_{\alpha \beta }^{a}\phi ^{a}(i)G_{\beta \alpha }(i,i)+{\rm Tr}\log G-{\rm Tr%
}[G_{0}^{-1}-G^{-1}]G  \label{bkphi} \\
&&-\frac{1}{2}{\rm Tr}\log \Pi +\frac{1}{2}{\rm Tr}[B^{-1}-\Pi ^{-1}]\Pi
+\Phi \lbrack G,\Pi ]
\end{eqnarray}
\begin{equation}
S[{\phi}]=\frac{1}{2}\int dxdy{\phi}_{a}(x)B_{ab}^{-1}(x-y){\phi}_{b}(y)
\label{sbar}
\end{equation}
$G,\Pi $ are the full fermion and boson Green function, and $\Phi \lbrack
G,\Pi ]$ is the sum of all 2--particle irreducible diagrams constructed
using the vertex (\ref{vertex}) 
with particle lines $G,\Pi $.

The extended DMFT equations are obtained by retaining only the local terms
i.e. $\Phi (G,\Pi )\approx \Phi _{EDMFT}$ =$\sum_{i}\Phi \lbrack G_{ii},\Pi
_{ii}]$. The importance of this truncation, is that its solution can be
formulated as a dynamical mean--field theory, that is one can introduce an
impurity model as in ref, the only difference is that now a Weiss field for
the electrons and for the bosons has to be introduced.

\begin{eqnarray}
G(i,i,i\omega _{n}) &=&[G_{0}^{-1}-{\frac{{\delta }\Phi }{{\delta G}}}%
]^{-1}(i,i) \\
\Pi (i,i,i\omega _{n}) &=&[B^{-1}+2{\frac{{\delta }\Phi }{{\delta \Pi }}}%
]^{-1}(i,i)  \nonumber \\
\phi _{b}(i) &=&ig\sum_{a}\sum_{\alpha \beta }B_{{}}^{ba}(i,j)G_{\alpha
\beta }^{{}}\Lambda _{\alpha \beta }^{a}  \label{edmfvac}
\end{eqnarray}

Our case of long range Coulomb interactions corresponds to $\Pi
_{0}(q,i\omega _{n})=V(q)$. The one--band case was considered in ref \onlinecite
{chitracoul}. It produces qualitatively new effects turning the Mott
transition from second order to first order. A full numerical solution of
this problem was only recently carried out for a model Hamiltonian \cite
{motome}, in a study of a Fermi Bose system.

E--DMFT treats both electrons and the collective excitations (spin and
charge fluctuations ) on a similar footing. The fermion (boson) propagators $%
G$ and ${\ \Pi }$ are expressed in terms of self energies ${\Sigma _{F}}%
(i\omega _{n})$ ${\Sigma _{B}}(i\omega _{n})$ which are taken to be momentum
independent, $G^{-1}(i\omega _{n},q)=i\omega _{n}-t_{q}{-\Sigma }%
_{F}(i\omega _{n}),{\Pi }(q,i\omega _{n})=-{\tilde{\Sigma _{B}}+B}(q,i\omega
_{n}).$The self--energies as well as other local quantities are computed
from the local action:

\begin{eqnarray}
S_{loc} &=&\int d\tau d\tau ^{\prime }\sum_{\sigma }c_{\sigma }^{+}(\tau
)G_{0\sigma }^{-1}(\tau -\tau ^{\prime })c_{\sigma }(\tau ^{\prime
})-\sum_{\sigma }\phi _{a}(\tau )\Pi _{ab}^{-1}(\tau -\tau ^{\prime })\phi
_{b}(\tau ^{\prime })+ \\
&&\int d\tau \sum_{\alpha \beta \gamma \delta }\Gamma ^{\alpha \beta \delta
\gamma }c_{i\alpha }^{+}c_{i\beta }^{+}c_{i\gamma }c_{i\delta } +
i\sum_{\alpha \beta }\int_{0}^{1/T}\phi ^{a}(i){\Lambda }_{a \alpha \beta
}^{a}c_{i\alpha }^{+}c_{i\beta }d\tau
\end{eqnarray}
and the parameters of the local action are determined by solving the E--DMFT
self consistency conditions

\[
\Pi _{0}^{-1}(i\omega _{n})=[\sum_{q}{\frac{1}{{\ -{\Sigma }_{B}^{{}}\{\Pi
_{0},G_{0}\}(i\omega _{n})+B(q,}i\omega _{n})}]}^{-1}+{{\Sigma }_{B}}\{\Pi
_{0},{G}_{0}\}(i\omega _{n}) 
\]

\begin{equation}
{G}_{0}^{-1}(i\omega _{n})=[\sum_{q}{\frac{1}{{\ i\omega _{n}-t_{q}-\Sigma }%
_{F}{\{\Pi _{0},G_{0}\}}}]}^{-1}+\Sigma _{F}\{\Pi _{0},{G}_{0}\}(i\omega
_{n})
\end{equation}
The advantage of this formulation is that it is naturally combined with
additional diagrams so that it resembles a natural extension of the GW
method. Hence we propose the approximation $\Phi_{GWU}= \Phi_{EMDFT} + {\Phi}%
^1$. ${\Phi}^1$ is the lowest order nonlocal correction (which is likely to
be much smaller that the ordinary Hubbard term).

\begin{equation}
\Phi ^{1}(G,D)=\frac{1}{2}\sum_{i\neq j}\sum_{ab}\sum_{\alpha \beta \gamma
\delta }G_{ij}^{\alpha \beta }G_{ij}^{\gamma \delta }\Lambda _{a}^{\alpha
\beta }\Lambda _{a}^{\gamma \delta }D_{ij}^{ab}  \label{nlocal}
\end{equation}
If indeed the local approximation is a good starting point, there should be
not much difference between a full self consistent treatement or a
perturbative lowest order treatment of the self energy derived from Eq. (\ref
{nlocal}).

\section{ Outlook}

It is important to stress in this work\ devoted to methods for treating
strongly correlated electron systems, the fruitful interplay between
many--body theory of model hamiltonians and the field of
electronic--structure calculations. The hard work in developing the
dynamical mean--field technique has paid off handsomely in insights into
real materials, and has already resulted in a major advance in
electronic--structure theory, i.e. in our ability to predict physical
properties of materials starting from first principles. Furthermore the
technology developed for the solutions of model hamiltonians has been
effectively transferred into the field of\ electronic structure. Simple
calculations which gave the first spectra of the Hubbard model, using
mapping onto impurity models\cite{antoine}, have been implemented with full
realism using the tetrahedron method\cite{poteryaev} . Simple functionals
which were developed to analyze the simplest possible
localization--delocalization transition\cite{landau,chitrafunc2}, \ have
been extended and implemented to yield total energies of solids\cite{pluprl}%
. Interdisciplinary exchanges are a two--way street, and many of the cluster
DMFT ideas which originate in advances in one--electron theory, hold great
promise for the use in model hamiltonians.\cite{CDMFT}. There is no doubt
that the study of simple systems has been essential \ in our quest to
control and understand complex materials

The DMFT\ techniques will continue to be used in a much wider range of
realistic problems, where correlation effects are prominent. In this
context\cite{vlad2} none of the basic ideas of the
local approximation, requires a translation invariant.\ In fact DMFT\ has
been applied to strongly disordered system which exhibit Anderson
localization and to surfaces \cite{nolting}. There are many physical reasons to believe
that correlation effects in alloys and interfaces are much stronger than in
the bulk of periodic solids DMFT\ can also be applied to finite systems such
as complex molecules or molecular clusters.

The next step in the development of DMFT as a realistic
electronic--structure method is its implementation in a molecular--dynamic
calculation, to predict structures without any a priori information. A
second\ important and largely unexplored direction is the study of systems
far from equilibrium. There have been substantial advances in understanding
single impurity Anderson model far from equilibrium in the context of
quantum dots, and this understanding will certainly allow major advances in
the treatment of solids when combined with the standard DMFT\
self--consistency condition.


\end{document}